\documentclass[12pt]{article}

\pdfoutput=1

\usepackage[sumlimits,intlimits,namelimits]{amsmath}
\usepackage{amssymb}
\usepackage[english]{babel}
\usepackage{bbm}
\usepackage{parskip}

\usepackage{graphicx}          %erlaubt Graphiken einzubinden

\newcommand{\AdS}{$\text{AdS}_{3}$}

\newcommand{\arctanh}{\text{arctanh}}

\newcommand{\bg}{\bar{g}}

\global\parskip 6pt
\newenvironment{definition}[1][Definition]{\begin{trivlist}
\item[\hskip \labelsep {\bfseries #1}]}{\end{trivlist}}

\newcommand{\qed}{\nobreak \ifvmode \relax \else\ifdim\lastskip<1.5em \hskip-\lastskip\hskip1.5em plus0em minus0.5em \fi \nobreak\vrule height0.75em width0.5em depth0.25em\fi}

\newcommand{\be}{\begin{eqnarray}}
\newcommand{\ee}{\end{eqnarray}}

\def\>{\rangle}
\def\<{\langle}

\usepackage{hyperref}% muss als letztes usepackage stehen

\begin{document}

%%%%%%%%%%%%%%%%%%%%%%%%%%%%%%%%%%%%%%%%%%%%%%%%%%%%%%%%%%%%%%%%%%%%%%%%
%Thu Dec  5 17:19:40 GMT 2002 Improved presentation. Clarified relation to linear stability and regularity of the metric
%%%%%%%%%%%%%%%%%%%%%%%%%%%%%%%%%%%%%%%%%%%%%%%%%%%%%%%%%%%%%%%%%%%%%%%%
\begin{titlepage}
\begin{flushright}
\hfill{LMU-ASC 21/13}
\\
\hfill{MPP-2013-109}
\end{flushright}
\vspace*{.5cm}
\begin{center}
{\Large{{\bf 
%Formation of 
Dynamical Black Holes \\[.5ex]
in $2+1$ Dimensions}}} \\[.5ex]
\vspace{1cm}
Mario Flory${}^{a,b}$\footnote{E-mail: mflory@mpp.mpg.de}
 and Ivo Sachs${}^{a}$\footnote{E-mail: ivo.sachs@physik.uni-muenchen.de}
\\
{\em ${}^{a}$Arnold Sommerfeld Center, 
Ludwig-Maximilians University,\\
Theresienstrasse 37, D-80333, Munich, 
Germany}
\\
{\em ${}^{b}$Max-Planck-Institut f\"ur Physik (Werner-Heisenberg-Institut),\\
F\"ohringer Ring 6, D-80805, Munich, 
Germany}\\
\end{center}
\vspace*{1cm}
\begin{abstract}
We investigate the global structure of a recently discovered simple exact, non-stationary solution of topologically massive and new massive gravity with the asymptotic charges of an undeformed BTZ black hole. We establish the existence of a timelike singularity in the causal structure of the spacetime even in the absence of angular momentum. The dynamical trapping and event horizons are determined and we investigate the evolution of the outer horizon showing that it may increase or decrease with time, depending on the value of the mass parameter. Finally, we test two proposals for dynamical entropy on this solution, one of them depending on the Kodama vector. In addition we show that the Kodama vector leads to the correct entropy for all stationary black holes in 2+1 dimensions.
\vspace*{.25cm}
%\noindent {PACS: 04.70Dy, 04.60.Kz, 11.25.Hf\ \ \ }
\end{abstract}
\vspace*{.25cm}
%December 2002
\end{titlepage}

%==============================================
%==========  Introduction  ===========================
%==============================================
\section{Introduction}
\label{sec:Intro}

Exact dynamical black hole solutions solutions in general relativity are notoriously hard to find. Known examples in four dimensions include the Oppenheimer-Snyder and the Vaidya solution (see e.g. \cite{OppenheimerS,Vaidya}). On the other hand we have a simple theory of gravity allowing for genuine black hole solutions, namely Einstein gravity in three dimensions with a negative cosmological constant. Although all classical solutions in this theory are locally equivalent to  three-dimensional anti-de Sitter space, a so-called BTZ black hole is obtained as a discrete quotient \cite{BTZ1,BTZ2} (see e.g. \cite{B1,Birmingham:2001dt} for a review). The price paid for this simplicity is that there is no propagating graviton in this theory. This, however, can be remedied by adding a higher derivative term to the action as was done in \textit{topologically massive gravity} (short TMG, see \cite{C1,C2}) and later in \textit{new massive gravity} (short NMG, see \cite{NMG,moreonNMG}).

Linear stability of the BTZ black hole has been established in \cite{D2} and its QNM spectrum in TMG was determined in \cite{D3}. Recently it was found however that away from the chiral point some of the linearized modes give rise to new exact non-stationary solutions to the TMG equations of motion with asymptotic charges equal to those of non-rotating BTZ black holes \cite{S1}.

In this article we provide a detailed description of the global structure of the resulting spacetime. We explicitly determine its trapping and event horizons, in particular, we show that the perturbed spacetime develops an inner horizon and that the formerly spacelike singularity of the BTZ BH is timelike. This may come as a surprise given the fact that the global charges are those of a non-rotating BTZ black hole \cite{S1} but is, of course, not in contradiction with Birkhoff's theorem. Still it is an interesting example of new phenomena that can arise once stationarity is abandoned. 

While the dynamical solution at hand was originally found to be a solution to TMG we will show that it also solves the equations of motion of NMG for suitably chosen parameters. This means, in particular, that we can apply a previously proposed definition for a dynamical entropy based on the dynamics of trapping horizons \cite{Hayward99,Hayward99b,Wu}. To do so, however, we have to revisit the definition of the Kodama vector for black holes in $2+1$ dimensions. As a by-product we show that the Kodama vector gives rise to the correct expression for the entropy for all stationary (i.e. including rotating) black holes in $2+1$ dimensions for NMG as well as generalizations thereof. The outcome of this procedure outlined below, however, leads to results in apparent contradiction with the second law and physical intuition when applied to such dynamical black holes.

The paper is organized as follows: In section \ref{sec:dynamical BH} we will discuss the dynamical black hole metric presented in \cite{S1}, in particular its event and trapping horizons. We will also show that this metric is a solution of NMG. In section \ref{sec:Kodama and Entropy} we will define the Kodama vector and elaborate on Hayward's approach to dynamical black hole entropy. We will apply this approach to the dynamical black holes in the sections \ref{sec:applied} and to the general stationary $2+1$ dimensional black hole in section \ref{sec:generalBH}. We end with a conclusion in section \ref{sec:Conclusion}. In appendix \ref{sec:NMG0} we will review the action of NMG and its properties such as unitarity, while in appendix \ref{sec:IW} we apply the definition of dynamical black hole entropy proposed by Iyer and Wald \cite{Wald50} to the dynamical black holes in the framework of TMG. In the entire work, we will use units in which $\hbar=c=k=1$ and the convention that spacetime indices in $d$ 
dimensions take values $\mu,\nu$ in \{0,1,...,$d-1$\}. We will also use $G_N=1/8$ if not stated otherwise.

%==============================================
%==========  dynamical BH  ===========================
%==============================================
\section{Dynamical Black Holes in three Dimensions}
\label{sec:dynamical BH}

To fix the notation let us first recall the line-element of the non-rotating BTZ-black hole with $M=1$ and $l=1$ (i.e. $\Lambda=-1$) \cite{BTZ1,BTZ2} which reads:
\begin{align}
ds^2=\bg_{\mu\nu}dx^{\mu}dx^{\nu}=-\sinh^2(\rho)dt^2+\cosh^2(\rho)d\phi^2+d\rho^2
\nonumber
\end{align}
Both in TMG and NMG the linearized gravitational perturbations are given by the solutions of an equation $\mathcal{D}^\mu{}_{\lambda} h_{\mu\nu}=0$ \cite{D1,correlators,BTZstability,Liu}, subject to suitable boundary conditions \cite{Henneaux:2009pw}. They can be classified  in terms of highest weight representation of the $sl(2,R)\times sl(2,R)$ isometry of  \AdS.  

%=================== The Metric ===================
\subsection{The Metric}
\label{sec:metric}

In \cite{S1} it was observed that adding a highest weight perturbation to the BTZ metric, i.e.
\begin{align}
g_{\mu\nu}=\underbrace{\left(
\begin{array}{ccc}
 -\sinh^2(\rho) & 0 &0 \\
 0 & \cosh^2(\rho) & 0 \\
0 & 0  & 1  \\
\end{array}
\right)}_{=\bg_{\mu\nu}}+
\underbrace{(e^{t}\sinh \rho)^{1 +\mu} 
\left(\begin{array}{ccc}
1&1&\frac{2}{\sinh(2 \rho)}\\
1&1&\frac{2}{\sinh(2\rho)}\\
\frac{2}{\sinh(2\rho)}&\frac{2}{\sinh(2\rho)}&\frac{4}{\sinh^2(2\rho)} 
\end{array}\right)}_{\equiv h_{\mu\nu}}
\label{sachsmetric}
\end{align}
yields a solution to the full nonlinear vacuum-equations of motion of TMG, $G_{\mu\nu}+\frac{1}{\mu}C_{\mu\nu}=0$ with $G_{\mu\nu}\equiv R_{\mu\nu}-\frac12\, g_{\mu\nu} R+\Lambda g_{\mu\nu}$, the \textit{Cotton tensor} $C_{\mu\nu}$ and the mass parameter $\mu$. Here, $h_{\mu\nu}$ is a metric that was first constructed as a solution to the \textit{linearized} equations on motion of TMG in \cite{D2,D3}. It should be noted that in the coordinate system we are using $x^0=t\in\ ]-\infty,+\infty[$, $x^2=\rho\in\ ]0,+\infty[$ and $x^1=\phi\in[0,2\pi[$ with $\phi\sim\phi+2\pi$. Obviously, the metric (\ref{sachsmetric}) has the structure ``background plus distortion". 

The Riemann- and Einstein-tensor of (\ref{sachsmetric}) were already calculated in \cite{S1} and read
\begin{align}
R_{\mu\nu}=\frac{R}{3}g_{\mu\nu}+\frac{\mu^2-1}{12}Rh_{\mu\nu}
\text{  ,  }
G_{\mu\nu}=\frac{1-\mu^2}{2}h_{\mu\nu}
\label{Etensor}
\end{align}
where we made use of the Ricci-scalar $R=\bar{R}=-6$ which for vacuum solutions is fixed by the trace of the equations of motion of TMG as a function of the AdS-radius. 

Having justified that (\ref{sachsmetric}) represents an exact vacuum-solution of the full non-linear equations of motion of TMG, we can ask: What kind of spacetime does this metric describe? In \cite{S1} the spacetime was already classified as a locally AdS pp-wave spacetime \cite{S1,Chow:2009km} of Petrov type N (see \cite{Chow:2009km}) and Kundt-CSI type (see \cite{Chakhad:2009em,Chow:2009vt}). Apart from these facts, as the metric (\ref{sachsmetric}) was derived from a black hole background and indeed asymptotes to this background in certain limits, we can already speculate that this metric might describe a dynamical black hole spacetime. 

%==========  global coord  ===========================
\subsection{Global coordinates}
\label{sec:global coord}

A good coordinate system should fulfill two requirements: Firstly, it should bring the line element to a simple form. Secondly, it should cover a large part of the spacetime. The coordinates that prove most useful for discussing the global structure of the metric $g_{\mu\nu}$ are defined by
\begin{align}
z=e^{-t}\frac{1}{\sinh(\rho)}\text{ , }R=e^{-2t}\coth^2(\rho)&\text{ , } y=t+\phi+\log\left(\tanh(\rho)\right)
\label{zyRcoord}
\end{align}
where we choose $x^0=z$, $x^1=y$ and $x^2=R$. %and $\epsilon^{zyR}=\frac{+1}{\sqrt{-g}}$
In these coordinates, the Killing vector $\partial_\phi$ is equal to $\partial_y$. %while $\xi=-2e^{-y}\partial_R$. 
The line element of the metric (\ref{sachsmetric}) takes the very simple form
\begin{align}
ds^2=\underbrace{\frac{1}{z^2}\left(dz^2+dydR+Rdy^2\right)}_{\bar g_{\mu\nu}dx^{\mu}dx^{\nu}}+\underbrace{\frac{1}{z^{1+\mu}}dy^2}_{h_{\mu\nu}dx^{\mu}dx^{\nu}}
\label{dsforzyR}
\end{align}  
Here, the second term on the right-hand side corresponds to the perturbation $h_{\mu\nu}$ while the first term corresponds to the background metric $\bg_{\mu\nu}$. Because of the factor $z^{-\mu}$, we need to restrict $z$ to positive values for general $\mu$. Apart from this we can set $y\in[0,2\pi[$ with $y\sim y+2\pi$ and $R\in]-\infty,+\infty[$. This coordinate system thus covers a much larger part of the spacetime than it was the case for the Schwarzschild-like coordinates used in (\ref{sachsmetric}). Therefore, we call these coordinates the \textit{global coordinates}.

The spacetime's structure is much easier to understand in these new coordinates. In \cite{BTZ2} it was pointed out that the singularity of BTZ black holes is not a curvature singularity but merely a singularity in the causal structure of the spacetime, implied by the presence of closed causal curves. In order to find out whether there is a similar singularity present in the family of metrics given by equation (\ref{sachsmetric}), we note that because of the periodicity in the coordinate $\phi$ and (\ref{zyRcoord}), the point $(z,R,y)$ is identified with the point $(z,R,y+2\pi)$. As closed causal curves therefore appear where $\partial_{y}$ is null or timelike, we have to restrict the \textit{physical part of the spacetime} to the region where $R>-z^{1-\mu}$, with the equation $R=-z^{1-\mu}$ determining the singularity. For the nonrotating BTZ background metric $\bg_{\mu\nu}$ the singularity is the hypersurface determined by the equation $R=0$. It can be shown that in the physical part of the spacetime, the 
coordinate $R$ has to decrease along every future pointing causal curve \cite{Thesis2}.

%==========  ev horizons  ===========================
\subsection{Event horizons}
\label{sec:ev horizons}

\begin{figure}[htb]                                 
\begin{center}                                      
	\includegraphics[width=0.45\linewidth]{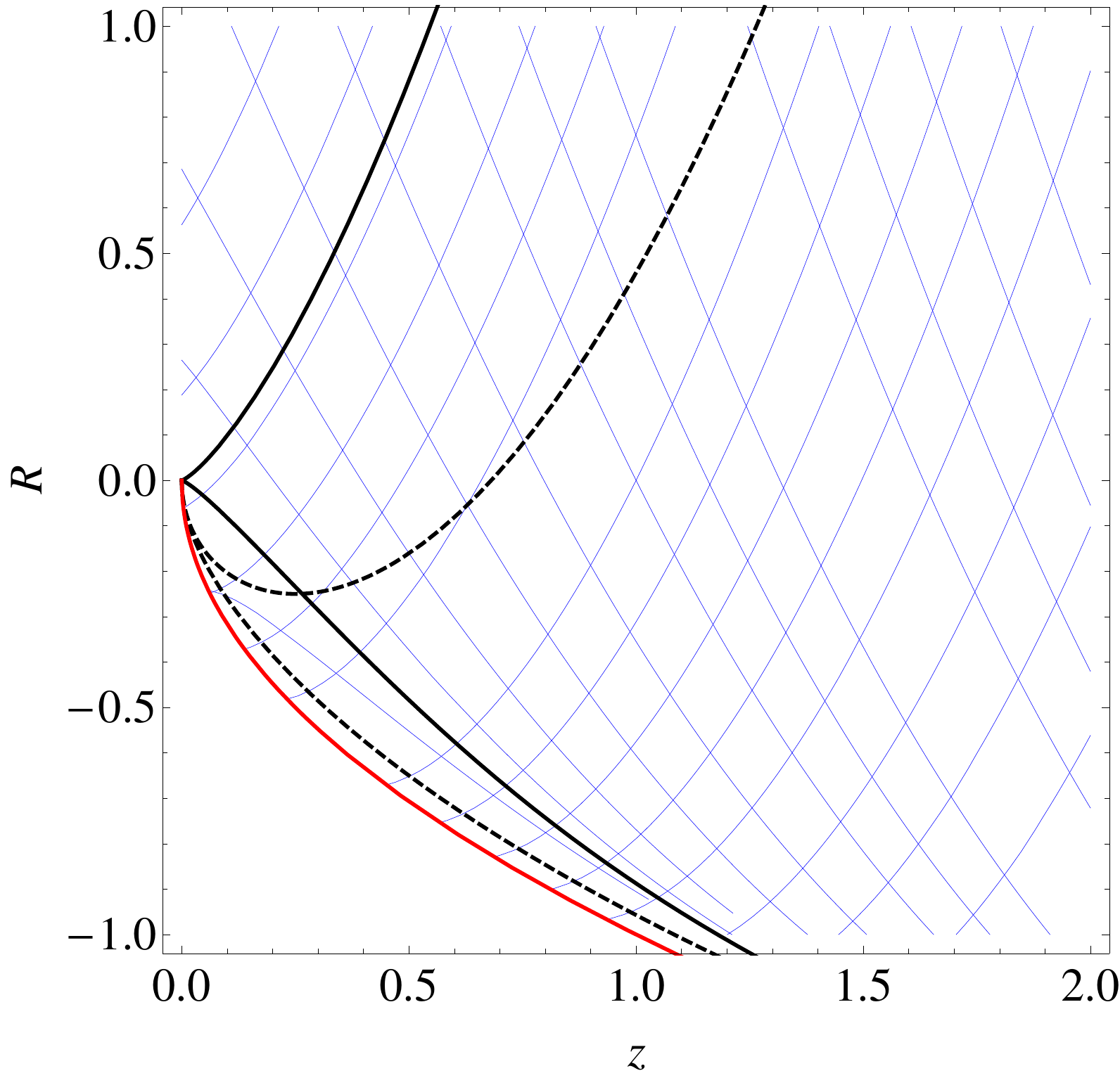}%{plots1.pdf}       
	\includegraphics[width=0.45\linewidth]{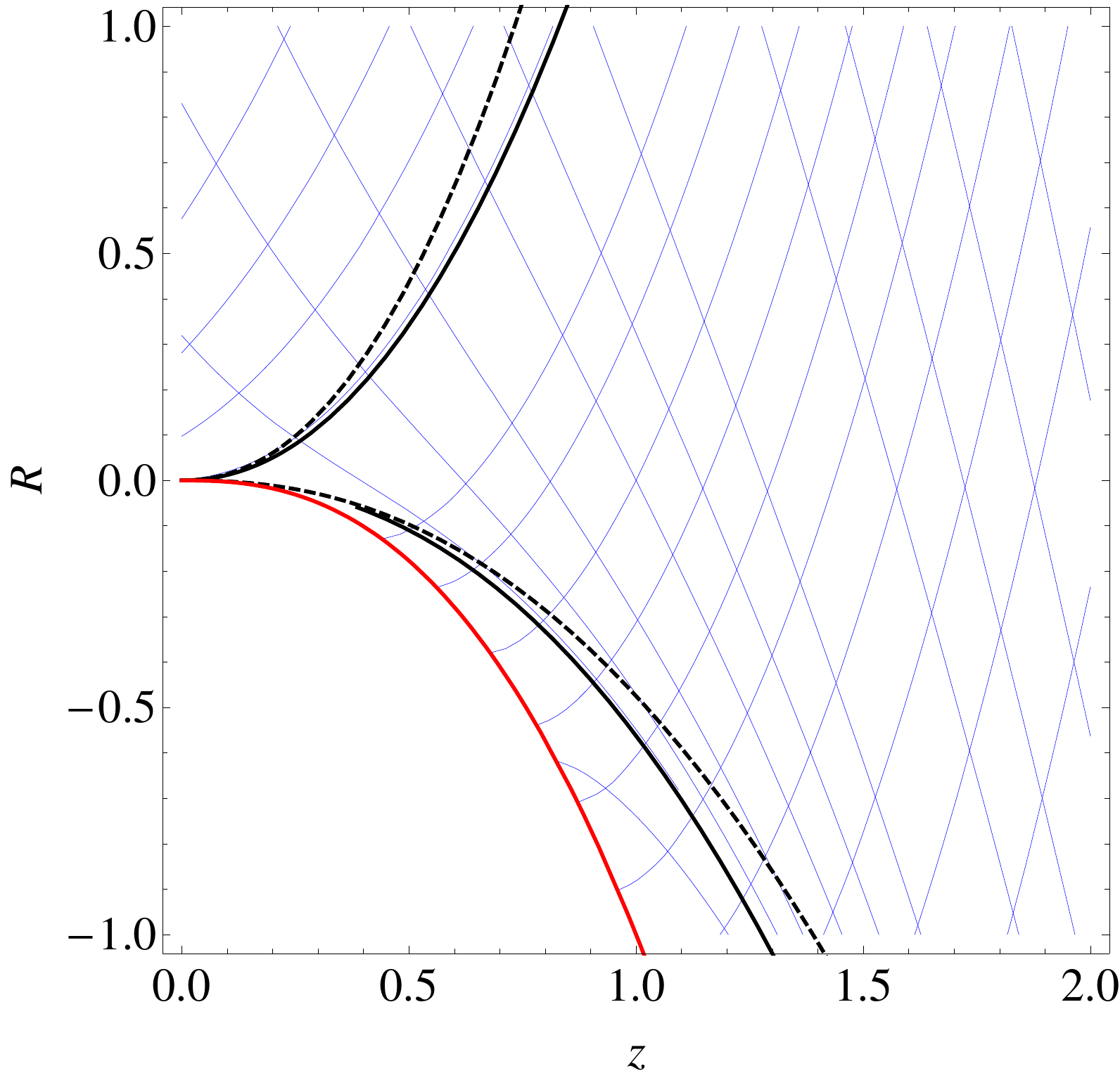}%{plots2.pdf}
\end{center}
\caption[]{Diagrams for $\mu=1/2$ on the left, and $\mu=-3/2$ on the right. Event horizons are depicted as solid black lines, trapping horizons as dashed black lines and the singularity as red line. The projections of several null geodesics of maximal slope in the $z$-$R$-plane are drawn as thin blue lines.}
\label{fig:overview}
\end{figure}

Having proven the existence of a singularity, it is natural to ask about the existence of event horizons. % as defined in section \ref{sec:evhorizons}. 
For simplicity, we will limit our investigation to the cases where $\mu\leq1$. Now, the global nature of the definition of event horizons becomes a problem, especially as the asymptotics of our spacetime at infinity are not necessarily AdS-like for general $\mu$. The limit $\rho\rightarrow\infty$ and $t=const.$ corresponds to $z\rightarrow 0$ and $R\rightarrow const.$ in global coordinates. We therefore adopt the viewpoint that in these coordinates, $z=0$, $R>\lim_{z\rightarrow 0}\left(-z^{1-\mu}\right)$ and $y$ being arbitrary describes ``infinity", and that the (outer) event horizon of the spacetime will be described by the boundary of its causal past. This ansatz is far from perfect, the possible problems of such an approach were discussed in \cite{Hubeny:2002pj}. We will nevertheless pursue this approach for three reasons: Firstly, it reproduces the correct event horizon in the cases $\mu=\pm1$ as we will see in sections \ref{sec:+1} and \ref{sec:-1}. Secondly, for $\mu<-1$ the asymptotics for $\rho\
rightarrow\infty$ are the same as in the BTZ-case as $(e^{t}\sinh(\rho))^{1+\mu}\rightarrow0$ in this limit. Thirdly, using this definition for $\mu<1$, in a spacetime diagram such as figure \ref{fig:overview} event and trapping horizons approach the same point $z=0=R$ in the limit $z\rightarrow0$.

In the following, we will show how to numerically determine the horizons. As the singularity contains a timelike direction for $\mu\neq1$, there will in general be an outer as well as an inner horizon.

Due to the definition of the outer and inner event horizons as boundaries between points from which a certain limit or hypersurface can be reached on causal curves\footnote{We define the inner event horizon to be the boundary between points in the physical part of the spacetime from which the singularity can be reached on past-pointing causal curves and such points in the physical part of the spacetime from which this is not possible.} and points from which this is not possible, the event horizons will be generated by null geodesics of maximal and minimal slope when projected down to the $z$-$R$-plane. % defined in (\ref{min/max}) and the following discussion in section \ref{sec:geodesics}. 
Therefore, the outer horizon is for $\mu<1$ defined to be the solution of the differential equation\footnote{For a discussion of geodesics (both analytical and numerical), lightcones and causal curves in the spacetime (\ref{sachsmetric}) see \cite{Thesis2}.}
\begin{align}
\frac{dR}{dz}=2\sqrt{\left(R+z^{1-\mu}\right)}\text{ with the initial condition $R(0)=0$.}
\label{outerevhorizon}
\end{align}
Similarly, the inner horizon is defined to be the solution of 
\begin{align}
\frac{dR}{dz}=-2\sqrt{\left(R+z^{1-\mu}\right)}\text{ with $R(0)=0$.}
\label{innerevhorizon}
\end{align}
Unfortunately, there is no closed-form expression for the solutions of these equations for $|\mu|\neq1$, but numerical solutions can be calculated. They are shown for $\mu=\frac{1}{2}$ and $\mu=-\frac{3}{2}$ together with the trapping horizons and the singularity in figure \ref{fig:overview}.

\begin{figure}[htbp]
\begin{center}
\includegraphics[width=0.5\linewidth]{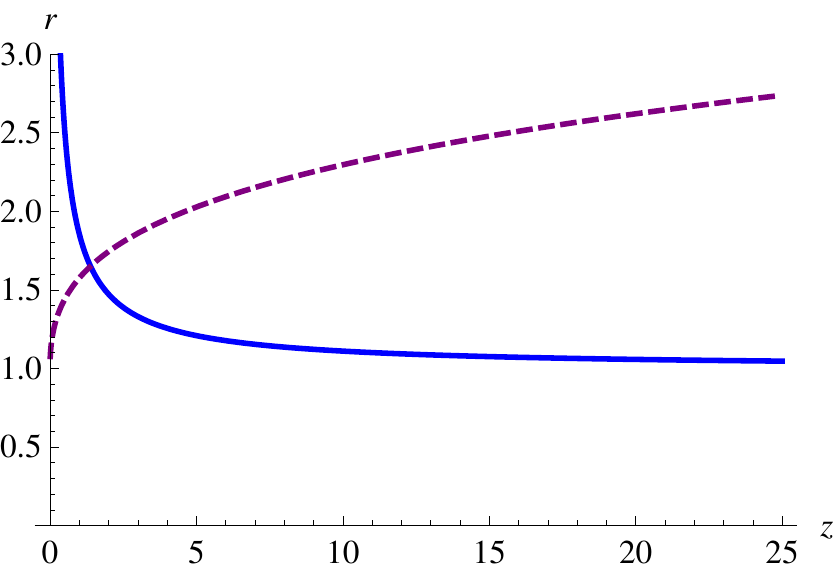} %besser kein Leerzeichen im Namen
\caption[]{Radii $r=\sqrt{g_{yy}}$ of the outer event horizon as a function of $z$ for $\mu=1/2$ (solid) and $\mu=-3/2$ (dashed). Smaller values of $z$ correspond to later times.}
\label{radii}
\end{center}
\end{figure}

Next we want to investigate the properties of the outer event horizon. The topology of the spacetime at hand is $\mathbb{R}^{2}\times S^{1}$ and in the physical part of the spacetime the radius of the compact dimension is (see section \ref{sec:Kodama and Entropy})
\begin{align}
r(z,R)=\sqrt{g_{yy}}=\frac{\sqrt{R+z^{1-\mu}}}{z}
\label{radius}
\end{align}
We can numerically compute the radius $r$ of the outer event horizon as a function of $z$. As the outer event horizon is always defined by a monotonous function $R(z)$, $z$ can be used as a measure of time instead of $R$, with large values of $z$ corresponding to early times and small values of $z$ corresponding to late times. Figure \ref{radii} shows the evolution of the outer event horizons as functions of $z$ for $\mu=1/2$ (solid) and $\mu=-3/2$ (dashed). While for $1>\mu>-1$ the horizon-circumference generally increases towards small $z$, it generally decreases for $-1>\mu$. 

%==========  Trapping hor  ===========================
\subsection{Trapping horizons}
\label{sec:trappinghorizons}

In this section we will recall the definition of \textit{trapping horizons}, which might be used as a definition of black hole boundaries instead of event horizons.

\begin{definition}\cite{Chakhad:2009em,Chow:2009vt,Frolov}:
In a $d$ dimensional spacetime ($d>2$), the \textit{expansion} $\theta$ of a null geodesic vector field $u^{\alpha}$ is defined to be 
\begin{align}
\theta=\frac{1}{d-2}u^{\alpha}_{;\alpha}
\label{expansion}
\end{align} 
\end{definition}

This definition allows us to mathematically formalize the trapping of a light ray in a strong gravitational field: 

\begin{definition}\cite{Hayward94,Battelle}:
Within a $d$ dimensional spacetime, a \textit{trapped surface} is a $(d-2)$ dimensional, closed, compact, spacelike surface $S$ such that for the expansions of the two families of future pointing null geodesics orthogonal to $S$, $\theta_{+}$ and $\theta_{-}$, $\theta_{+}\theta_{-}>0$ holds everywhere on $S$. The surface is called \textit{past trapped} or \textit{anti trapped} when $\theta_{\pm}>0$ everywhere on $S$, and \textit{future trapped} when $\theta_{\pm}<0$ everywhere on $S$. 
\end{definition}

Past trapped surfaces are typical for the interiors of white holes while future trapped surfaces are typical for black hole interiors. In order to describe black and white hole \textit{boundaries}, the previous definition has to be refined in the following way:

\begin{definition}\cite{Hayward94}: 
A \textit{marginal surface} is a $(d-2)$ dimensional, closed, compact spacelike surface $S$ such that either $\theta_{+}$ or $\theta_{-}$ (but not both) vanish on $S$.
\end{definition}

\begin{definition} \cite{Hayward94,Hayward94b}:
A \textit{trapping horizon} $\bar{H}$ is the closure of a $(d-1)$ dimensional surface $H$ foliated by marginal surfaces with $\theta_{a}\neq 0$ and $\mathcal{L}_{a}\theta_{b}\neq0$ everywhere on $H$. Here, we use the notation $a\neq b$, $ a,b \in \{+,-\}$ and $\mathcal{L}_{\pm}$ denotes the Lie-derivative with respect to the out- or ingoing null geodesic vector field orthogonal to the marginal surfaces.   
\end{definition}
%\todo{(explain outer vs. inner, future vs. past??? nur wenn ich es spaeter brauche!!!)}

As the trapping horizons can be calculated as the hypersurfaces where the Kodama vector is a null vector (see section \ref{sec:Kodama and Entropy} or \cite{Hayward96}), we will not give a detailed derivation here. Instead, we will merely state the results:
\begin{align}
R_{+}(z)=\frac{1}{2} z^{-2 \mu} \left((-\mu-1) z^{\mu+1}+z^{2 \mu+2}+\sqrt{z^{3 \mu+3} \left(z^{\mu+1}-2 \mu+2\right)}\right)
\label{outertrapping}
\\
R_{-}(z)=\frac{1}{2} z^{-2 \mu} \left((-\mu-1) z^{\mu+1}+z^{2 \mu+2}-\sqrt{z^{3 \mu+3} \left(z^{\mu+1}-2 \mu+2\right)}\right)
\label{innertrapping}
\end{align}
Let us shortly discuss the properties of the hypersurfaces described by these curves.

For $\mu<1$ it is easy to show that both $R_{+}(z)\rightarrow0$ and $R_{-}(z)\rightarrow0$ in the limit $z\rightarrow0$. This means that in global coordinates, both trapping horizons, event horizons and the singularity meet at $R=0=z$. Furthermore, one can show that $R_{-}(z)\geq-z^{1-\mu}$ for any $\mu$ with equality for $\mu<1$ only for $z=0$ or the limit $z\rightarrow+\infty$, which means that both trapping horizons (as $R_{+}(z)\geq R_{-}(z)$) will always be in the physical part of the spacetime.

Another interesting feature is that while for $\mu\leq-1$, $R_{+}(z)$ is a monotonous function of $z$, for $|\mu|<1$ the function $R_{+}(z)$ initially decreases, attains a minimum and then increases again with $z$. This ``bow" of the outer trapping horizon is quite unphysical if we want the trapping horizon to be a description of the black hole boundary. This means that there are points in the spacetime which are outside of the outer trapping horizon but which have a coordinate $R<0$, and from which it is not possible to escape the singularity. See figure \ref{fig:overview} for a plot of the trapping horizons for $\mu=\frac{1}{2}$ and $\mu=-\frac{3}{2}$.

It is possible to calculate the determinant $\mathfrak{g}$ of the induced metric on the trapping horizon, which contains one important physical information: its sign. For $\mu=\pm1$ it follows that $\mathfrak{g}(z)=0$ which means that the outer trapping horizon is a null-surface in these cases. For $\mu<-1$ we find $\mathfrak{g}(z)<0$ for any $z$ which means that in these cases the outer trapping horizon is a timelike hypersurface with signature $(-1,+1)$. For $|\mu|<1$ nevertheless, $\mathfrak{g}(z)<0$ for small $z$ and $\mathfrak{g}(z)>0$ for large $z$, indicating that due to the bow discussed above and shown in figure \ref{fig:overview}, the outer trapping horizon switches from a spacelike to a timelike hypersurface for some value of $z$. We called the trapping horizon (\ref{outertrapping}) ``outer" as it resembles the outer event horizon for $\mu\leq1$. Hayward \cite{Hayward94} used the deviating terminology that a trapping horizon is \textit{outer} when the expansion of the family of null geodesics that 
vanishes on the horizon \textit{shrinks} while passing through the horizon following the other family of null geodesics (with non-vanishing expansion) and \textit{inner} when it \textit{grows}. In this sense, what we called the outer trapping horizon changes from being an outer trapping horizon to being an inner trapping horizon when $z\rightarrow0$.  

%==========================================
%=============  NMG  =========================
%==========================================
\subsection{Dynamical black holes in NMG}
\label{sec:NMG}

In \cite{S1} it was shown that the metric (\ref{sachsmetric}) is a solution to the full non-linear vacuum equations of TMG. By comparing the linearized equations of motion for massive modes of TMG and NMG (see \cite{D1} for TMG and \cite{correlators,BTZstability,Liu} for NMG), it can be seen that a linear solution of TMG will also be a linear solution of NMG if we set $m^2=\mu^2-1/2$\footnote{This is also what was found in \cite{Ahmedov:2010em} for a solution of TMG of Petrov type N to be a solution of NMG.}. We can now ask whether we will have the same effect for NMG as for TMG, i.e. whether the metric (\ref{sachsmetric}) is also a solution of the full non-linear equations of motion (\ref{EOMofNMG}) of NMG (see appendix \ref{sec:NMG0}). In order to answer this question, it is advisable to first consider the trace of the equations of motion of NMG, (\ref{TraceofNMG}). Making use of $R=-6$ and (\ref{Etensor}) it is easy to find $K=R_{\mu\nu}R^{\mu\nu}-\frac{3}{8}R^2=-\frac{3}{2}$ independently of $\mu$, which 
is also the case for the background metric $\bg_{\mu\nu}$. (\ref{TraceofNMG}) then reads
\begin{align}
&6+6\lambda+\frac{3}{2m^2}=0
\nonumber
\\
\Rightarrow &\lambda = \frac{-4 m^2-1}{4 m^2}\text{  or  } m^2=-\frac{1}{4 (\lambda +1)}
\label{conditionsforNMG}
\end{align}
which is equivalent to (\ref{AdSinNMG}) for $\Lambda=-1$. Inserting now (\ref{sachsmetric}) in (\ref{EOMofNMG}) using $m^2=-\frac{1}{4 (\lambda +1)}$ yields
\begin{align}
R_{\mu\nu}-\frac{1}{2} g_{\mu\nu} R +\lambda g_{\mu\nu}-\frac{1}{2m^2}K_{\mu\nu}=\frac{1}{2} \left(1-\mu^2\right) \left( 4 (\lambda +1) \mu^2-2 \lambda-1\right)h_{\mu\nu}
\label{NMGEOM}
\end{align}
The right-hand side is obviously zero for $\mu=\pm1$ and $\mu=\pm\sqrt{\frac{2 \lambda +1}{4 \lambda +4}}$ or equivalently $\lambda=\frac{1-4 \mu^2}{2 \left(2 \mu^2-1\right)}$. This means that for the correct choices of the parameters $\mu$, $m^2$ and $\lambda$, (\ref{sachsmetric}) is also a solution to NMG. Inserting this relation between $\mu$ and $\lambda$ into the relation (\ref{conditionsforNMG}) yields the expression $m^2=\mu^2-\frac{1}{2}$ which is exactly the condition that we where expecting from the comparison between the linearized equations of motion of TMG and NMG above.

Exact solutions of NMG have been studied extensively in the past, and the conditions under which certain solutions of TMG are also solutions of NMG have been investigated for example in \cite{Chakhad:2009em} and  \cite{Ahmedov:2010em}. It would be interesting to investigate how the solution (\ref{sachsmetric}) fits into the general families of exact solutions presented in \cite{Ahmedov:2010uk} and \cite{AdSwaves}. This will be left to future research. It seems nevertheless that in \cite{S1} and this work the metric (\ref{sachsmetric}) was first investigated as describing a dynamical black hole. 

%==========================================
%=============  Kodama  =========================
%==========================================
\section{The Kodama vector and dynamical entropy}
\label{sec:Kodama and Entropy}

%=============  Intro  =========================
%\subsection{Introduction}
%\label{sec:Kodama intro}

%\todo{check all this stuff???!!!}
In 1980, Kodama \cite{Kodama} investigated four dimensional black hole spacetimes with spherical symmetry. He found that in this case a vector field can be defined which coincides with the timelike Killing vector in the stationary case up to normalization and thereby offers a possible generalization of the timelike Killing vector to dynamic spacetimes \cite{Hayward99,Hayward99b,Wu,Hayward96,Kodama,Hayward97}. We will now present a generalization of this approach to dimensions $d\geq3$\footnote{While the sources used in this work \cite{Hayward99,Hayward99b,Wu,Hayward96,Kodama,Hayward97} restrict their discussion to four dimensional spacetimes, a generalization of the Kodama vector to $d$ dimensions has been discussed in \cite{Maeda:2007uu,Maeda:2010bu}. Nevertheless these authors assume that the coordinate system can be brought to a block diagonal (or \textit{warped product}) form $ds^2=g_{\alpha\beta}dy^{\alpha}dy^{\beta}+r^2(y)\gamma_{ij}(z)dz^{i}dz^{j}$ (with $\alpha,\beta\in\{0,1\}$, $i,j\in\{2,...,d-1\}$)
, which is not necessarily possible for a three dimensional metric with rotational symmetry.}:

Suppose we have a $d$ dimensional spacetime $\mathcal{M}$ which has the symmetry of a $(d-2)$ dimensional (hyper-)sphere $S^{d-2}$ with all corresponding Killing vectors being spacelike. Starting from any point $\mathcal{P}$ in the spacetime and following the flows of the Killing vectors of this symmetry will generate a $(d-2)$-sphere as spacelike submanifold. This sphere will be a geometrical invariant, therefore its $(d-2)$-volume $\mathcal{V}$ and its thereby defined aerial radius $r=\left(\mathcal{V}\frac{\Gamma\left((d-1)/2)\right)}{2\pi^{(d-1)/2}}\right)^{\frac{1}{d-2}}$ will be coordinate invariant scalar quantities defined at every point in the spacetime. Because of this, $\nabla_{\mu}r=\partial_{\mu}r$ will fix a well-defined one-form. This one-form can now be contracted with the binormal\footnote{We define the binormal to a closed spacelike surface $S$ as $\epsilon^{\mu\nu}=l^{\mu}n^{\nu}-l^{\mu}n^{\nu}$ where $l^{\mu}$ is the ingoing and $n^{\mu}$ is the outgoing null vector field orthogonal to 
$S$ with $l^{\mu}n_{\mu}=-1$ \cite{Tachikawa}. It obviously follows $\epsilon_{\mu\nu}\epsilon^{\mu\nu}=-2$.} $\epsilon^{\mu\nu}$ of the 2 dimensional space orthogonal to the $(d-2)$-sphere at $\mathcal{P}$ to yield the Kodama vector
\begin{align}
k^{\mu}=\epsilon^{\mu\nu}\partial_{\nu}r
 \label{Kodama}
\end{align}
as it was defined for $d=4$ in \cite{Hayward99,Wu}. In the case of our spacetime (\ref{sachsmetric}), we find for general $\mu$:
\begin{align}
k^{\mu}=
\left(
\begin{array}{c}
 -z \\
 \frac{(\mu-1) z}{2 \left(R z^\mu+z\right)}+1 \\
 -(\mu+1) z^{1-\mu}-2 R \\
\end{array}
\right)
\label{Kodama2}
\end{align}

The geometrical meaning of the Kodama vector field is that it is tangent to constant $r$-hypersurfaces, as obviously $k^{\mu}\partial_{\mu}r=\epsilon^{\mu\nu}\partial_{\nu}r\partial_{\mu}r=0$ due to the antisymmetry of $\epsilon^{\mu\nu}$ \cite{Hayward96}. Therefore, the Kodama vector field is spacelike in trapped regions, null on trapping horizons and timelike otherwise \cite{Hayward96}, making it easy to calculate trapping horizons when the Kodama vector field is known. 

Furthermore, in four dimensions it can be proven that Kodama and Killing vector agree in stationary, spherically symmetric spacetimes if the vector fields $k^{\mu}$ and $g^{\mu\nu}\partial_{\nu}r$ commute \cite{Hayward97}.

In \cite{Wald48,Kang49,Wald50} it was shown how the entropy of a stationary black hole can be calculated via the Noether charge associated with a certain Killing field. In dynamic spacetimes such a Killing vector field does not exist, but it was suggested by Hayward and others \cite{Hayward99,Hayward99b,Wu} that one could use the Kodama vector defined above as a generalization of the Killing vector to dynamic spacetimes, and thereby assign entropy to the trapping horizon of a dynamical black hole via a Noether charge approach.

First, one has to define the dynamical surface gravity $\kappa$ associated with the trapping horizon via \cite{Hayward99,Wu}
\begin{align}
\kappa=\frac{1}{2}\epsilon^{\alpha\beta}\nabla_{\alpha}k_{\beta}
\label{kappa}
\end{align}
For a theory of the form $S=\frac{1}{16\pi G_N}\int dx^{d}\sqrt{-g}L(g_{\mu\nu},R_{\alpha\beta\gamma\delta})$ the entropy of a spacelike slice $\Sigma'$ of the trapping horizon is then proposed to be \cite{Wu}
\begin{align}
\mathcal{S}=\frac{1}{16 G_N\kappa}\int_{\Sigma'}Q^{\mu\nu}\epsilon_{\mu\nu}\sqrt{\gamma}dy^{d-2}
\label{S_Hayward}
\end{align}
where again $\epsilon_{\mu\nu}$ is the binormal defined above and $\sqrt{\gamma}dy^{d-2}$ is the volume element on $\Sigma'$. $Q^{\mu\nu}$ are the components of the Noether charge $(d-2)$-form corresponding to $k^{\mu}$ given by \cite{Wu,37,38}
\begin{align}
Q^{\alpha\beta}=2\left[X^{\alpha\beta\mu\nu}\nabla_{\mu}k_{\nu}-2k_{\nu}\nabla_{\mu}X^{\alpha\beta\mu\nu}\right]
\label{Qcomponents}
\end{align}
with $X^{\alpha\beta\gamma\delta}\equiv\frac{\partial L}{\partial R_{\alpha\beta\gamma\delta}}$. For NMG one finds:
\begin{align}
Q^{\alpha\beta}=&\left(\frac{1}{2}+\frac{3}{8m^2}R\right)\left(\nabla^{\alpha}k^{\beta}-\nabla^{\beta}k^{\alpha}\right)
\nonumber
\\
&-\frac{1}{2m^2}\left(\nabla^{\alpha}k^{\nu}R^{\beta}_{\nu}-\nabla^{\mu}k^{\alpha}R^{\beta}_{\mu}-\nabla^{\beta}k^{\nu}R^{\alpha}_{\nu}+\nabla^{\mu}k^{\beta}R^{\alpha}_{\mu}\right)
\label{Noether components}
\\
&+\frac{1}{m^2}\left(k_{\nu}\nabla^{\alpha}R^{\beta\nu}-k^{\alpha}\nabla_{\mu}R^{\beta\mu}-k_{\nu}\nabla^{\beta}R^{\alpha\nu}+k^{\beta}\nabla_{\mu}R^{\alpha\mu}\right)
\nonumber
\end{align}
This proposal to dynamical entropy will be called \textit{Hayward's approach}. We will use this approach in the following subsections to calculate the dynamical entropy of the black holes given by (\ref{sachsmetric}) in the framework of NMG\footnote{There are formulas similar to (\ref{Qcomponents}) for TMG \cite{Bonora}, but evaluating these on a dynamical trapping horizon does not give a coordinate invariant result. Therefore, we will not present any results of Hayward's approach applied to dynamical black holes in the framework of TMG}.

\section{Hayward's approach applied to the dynamical black holes}
\label{sec:applied}

%=============  +1  =========================
\subsection{$\mu=+1$}
\label{sec:+1}

From (\ref{Etensor}) it follows that the metric (\ref{sachsmetric}) is not only a vacuum solution of TMG and NMG, but also a solution of ordinary Einstein gravity in the chiral cases $\mu=\pm1$. These special cases shall be investigated in more details in this and the following subsection before turning to the general case.

It can easily be seen that in the case $\mu=+1$ the line element (\ref{dsforzyR}) is equivalent to the line element $d\bar{s}^2=\bg_{\mu\nu}dx^{\mu}dx^{\nu}$ of the undisturbed BTZ black hole which can be verified by a simple coordinate shift $R'=R+1$. The Kodama vector (\ref{Kodama2}) can then be transformed to the Schwarzschild-like coordinates used in (\ref{sachsmetric}) and one finds for $\mu=+1$ that $k^{\mu}\partial_{\mu}=\partial_{t}$, i.e. that the Kodama vector equals the timelike Killing vector field in this static case, as expected. Consequently, Hayward's approach to black hole entropy will by definition yield the correct values for entropy and $\kappa$ in this case.  

%=============  -1  =========================
\subsection{$\mu=-1$}
\label{sec:-1}
Before moving on to the investigation of the case $\mu=-1$, we will comment on a detail of the metric (\ref{sachsmetric}) that was not addressed so far. In \cite{S1} it was described how the solution $h_{\mu\nu}$ of the linearized equations of motion of TMG around the background $\bar{g}_{\mu\nu}$ describes the metric $g_{\mu\nu}=\bar{g}_{\mu\nu}+h_{\mu\nu}$ which is a solution to the full equations of motion of TMG. But solutions to the linearized equations of motion can have arbitrary prefactors, and in general, we could have multiplied $h_{\mu\nu}$ with an arbitrary prefactor $\Xi$. Even if we had chosen to do so in section \ref{sec:NMG}, this would not have affected the fact that (\ref{sachsmetric}) fulfills the non-linear equations of motion.

For (\ref{sachsmetric}), such a prefactor $\Xi$ can obviously always be absorbed \textit{up to sign} by a shift in the coordinate $t$, except for the case where $\mu=-1$, as there the prefactor of $h_{\mu\nu}$ becomes $(e^{t}\sinh \rho)^{1 +\mu} =1$. Therefore, $h_{\mu\nu}$ actually describes \textit{two distinct}\footnote{Of course, the choice $\Xi=0$ would lead to the trivial solution $g^{0}_{\mu\nu}=\bar{g}_{\mu\nu}$. We nevertheless do not explicitly exclude the possibility $\Xi=0$ as for the continuum of solutions at $\mu=-1$ this value will be important, too.} one-parameter families of exact solutions of TMG, $g_{\mu\nu}(\mu)=\bar{g}_{\mu\nu}+h_{\mu\nu}(\mu)$ and $g'_{\mu\nu}(\mu)=\bg_{\mu\nu}-h_{\mu\nu}(\mu)$, which at the point $\mu=-1$ are connected by a continuum of non-isometric solutions $g_{\mu\nu}^{\Xi}=\bar{g}_{\mu\nu}+\Xi h_{\mu\nu}(-1)$.

Let us now come back to the metric $g_{\mu\nu}$ with $\mu=-1$ and $\Xi=1$. In this case, the singularity still contains a timelike direction and there are still two horizons, an outer and an inner one as discussed in sections \ref{sec:ev horizons} and \ref{sec:trappinghorizons}. The metric can therefore not be globally equivalent to the background metric $\bg_{\mu\nu}$ as was the case for $\mu=+1$, but might describe a rotating black hole with parameters $M\neq1$, $J\neq0$. Motivated by these considerations, we can now search for a coordinate transformation that maps the metric (see (\ref{sachsmetric}))
\begin{align}
g_{\mu\nu}^{\Xi}=
\left(
\begin{array}{ccc}
 -\sinh^2(\rho) & 0 &0 \\
 0 & \cosh^2(\rho) & 0 \\
0 & 0  & 1  \\
\end{array}
\right)
+\Xi
\left(\begin{array}{ccc}
1&1&\frac{2}{\sinh(2 \rho)}\\
1&1&\frac{2}{\sinh(2\rho)}\\
\frac{2}{\sinh(2\rho)}&\frac{2}{\sinh(2\rho)}&\frac{4}{\sinh^2(2\rho)} 
\end{array}\right)
\label{startwith}
\end{align}
to the BTZ metric for $l=1$ (with $x^0=t'$, $x^1=\phi'$, $x^2=r$):
\begin{align}
g_{BTZ\mu\nu}=
\left(
\begin{array}{ccc}
 M-r^2 & -\frac{J}{2} & 0 \\
 -\frac{J}{2} & r^2 & 0 \\
 0 & 0 & \frac{1}{\frac{J^2}{4 r^2}+r^2-M} \\
\end{array}
\right)
\label{targetmetric}
\end{align}
with parameters $M$ and $J$ that will certainly depend on $\Xi$. Such a coordinate transformation can easily be found and reads:
\begin{align}
t&=t'+\frac{1}{4} \left(-2 \log \left(r^2-1-\Xi \right)+\frac{2 \arctanh\left(\frac{2 \Xi -2 r^2+1}{\sqrt{4 \Xi +1}}\right)}{\sqrt{4 \Xi+1}}+\log \left(\Xi ^2+r^4-(2 \Xi +1) r^2\right)\right)
\nonumber
\\
\phi&=\phi'+\frac{1}{4} \left(-\log \left(\left(r^2-\Xi \right)^2-r^2\right)+2 \log \left(r^2-\Xi\right)-\frac{2 \arctanh\left(\frac{-2 \Xi +2 r^2-1}{\sqrt{4 \Xi +1}}\right)}{\sqrt{4 \Xi +1}}\right)
\nonumber
\\
\rho& =\cosh ^{-1}\left(\sqrt{r^2-\Xi }\right)
\nonumber
\end{align}
This transformation maps the metric (\ref{startwith}) to the metric (\ref{targetmetric}) with parameters $M=1+2\Xi$ and $J=-2\Xi$\footnote{This means that we singled a one parameter ($\Xi$) family out of the two parameter ($M,J$) space of BTZ black holes for $l=1$. These black holes are exactly those with an entropy $\mathcal{S}=\frac{\pi}{2G_N}$ in the framework of TMG with $\mu=-1$, $l=1$.}.
%It should be noted that the BTZ metric with $M$ and $J$ can always be transformed to the BTZ metric with $M'=M$ and $J'=-J$ by a reversal of time, $t\rightarrow t'=-t$. 
Above we restricted ourselves to transformations that do not reverse time. It should also be noted that in the form used above this coordinate transformation is only valid for $r>\frac{1}{2} \left(\sqrt{4 \Xi +1}+1\right)$, and this lower bound can indeed be shown to be the radius $r_{+}$ of the outer event horizon of the black hole with $M=1+2\Xi$, $J=-2\Xi$. Interestingly, the cosmic censorship bound $Ml\geq |J|$ is only fulfilled for $\Xi\geq-\frac{1}{4}$. %\todo{here again G=1/8}

As the metric (\ref{startwith}) describes a rotating but stationary black hole, Hayward's approach reproduces the correct entropy as shown in section \ref{sec:generalBH}.

%=============  <1  =========================
\subsection{$\mu<1$, $\mu\neq-1$}
\label{sec:mu<1}

\begin{figure}[Hhtbp]                                 
\begin{center}
\includegraphics[width=0.3\linewidth]{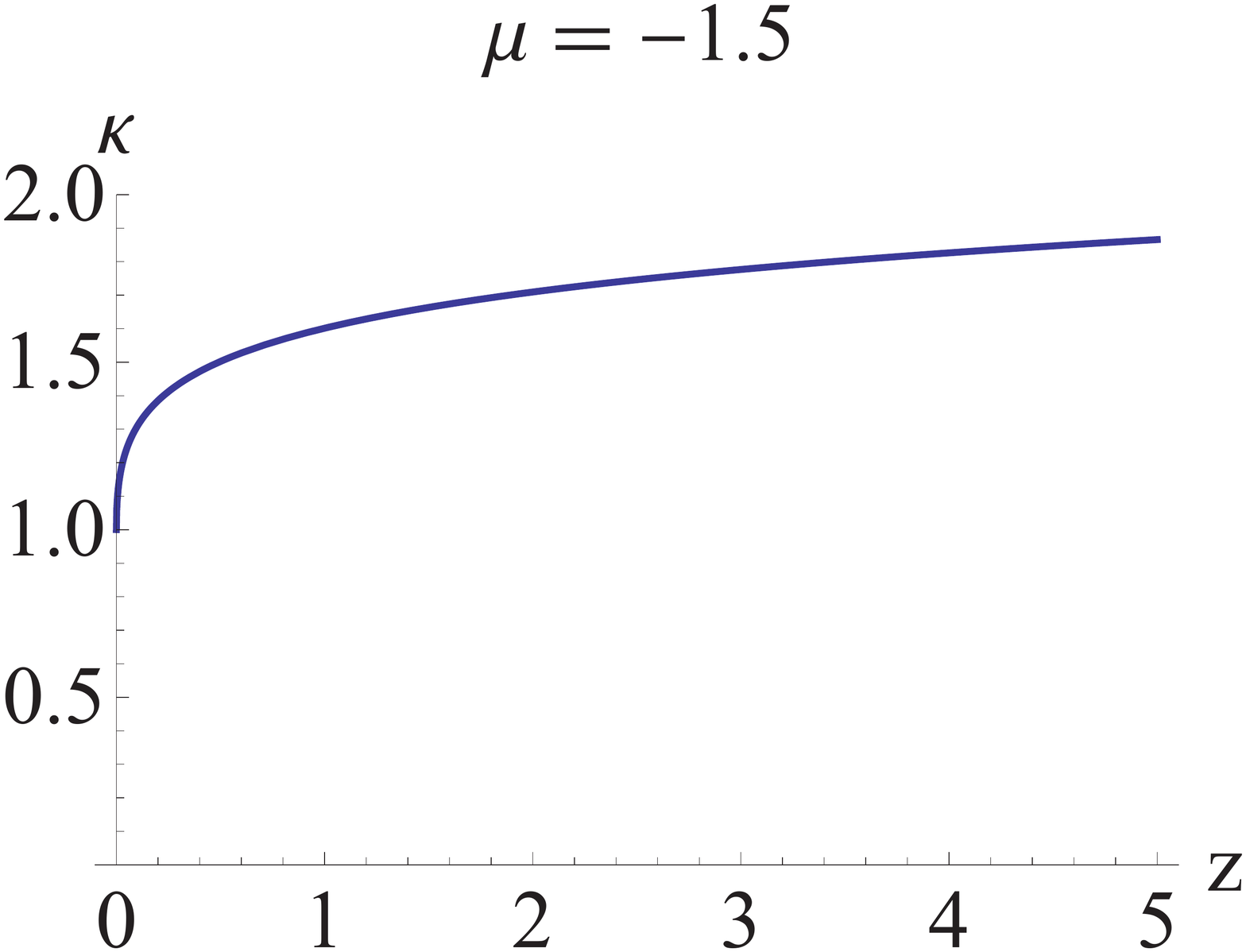}  
\includegraphics[width=0.3\linewidth]{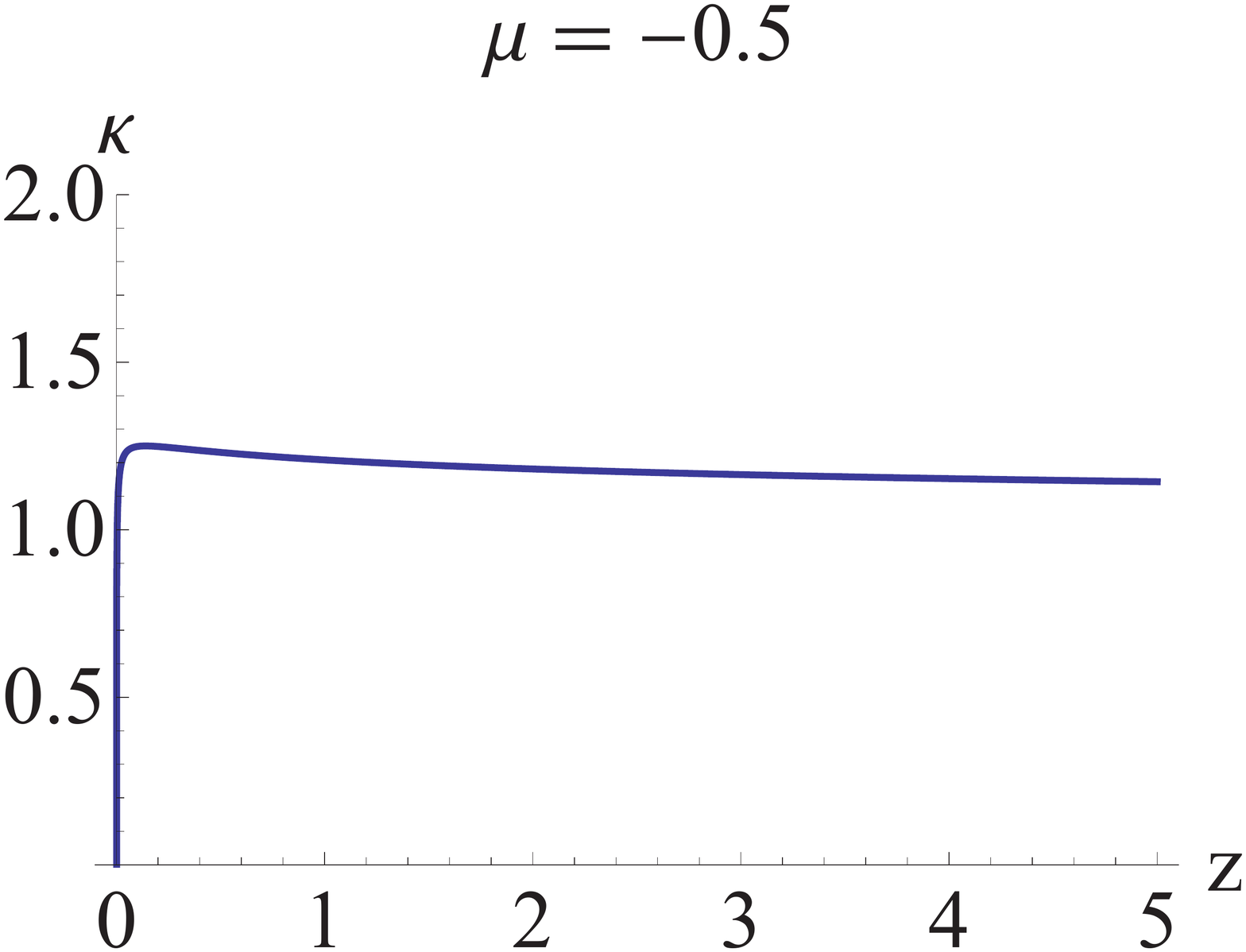}  
\includegraphics[width=0.3\linewidth]{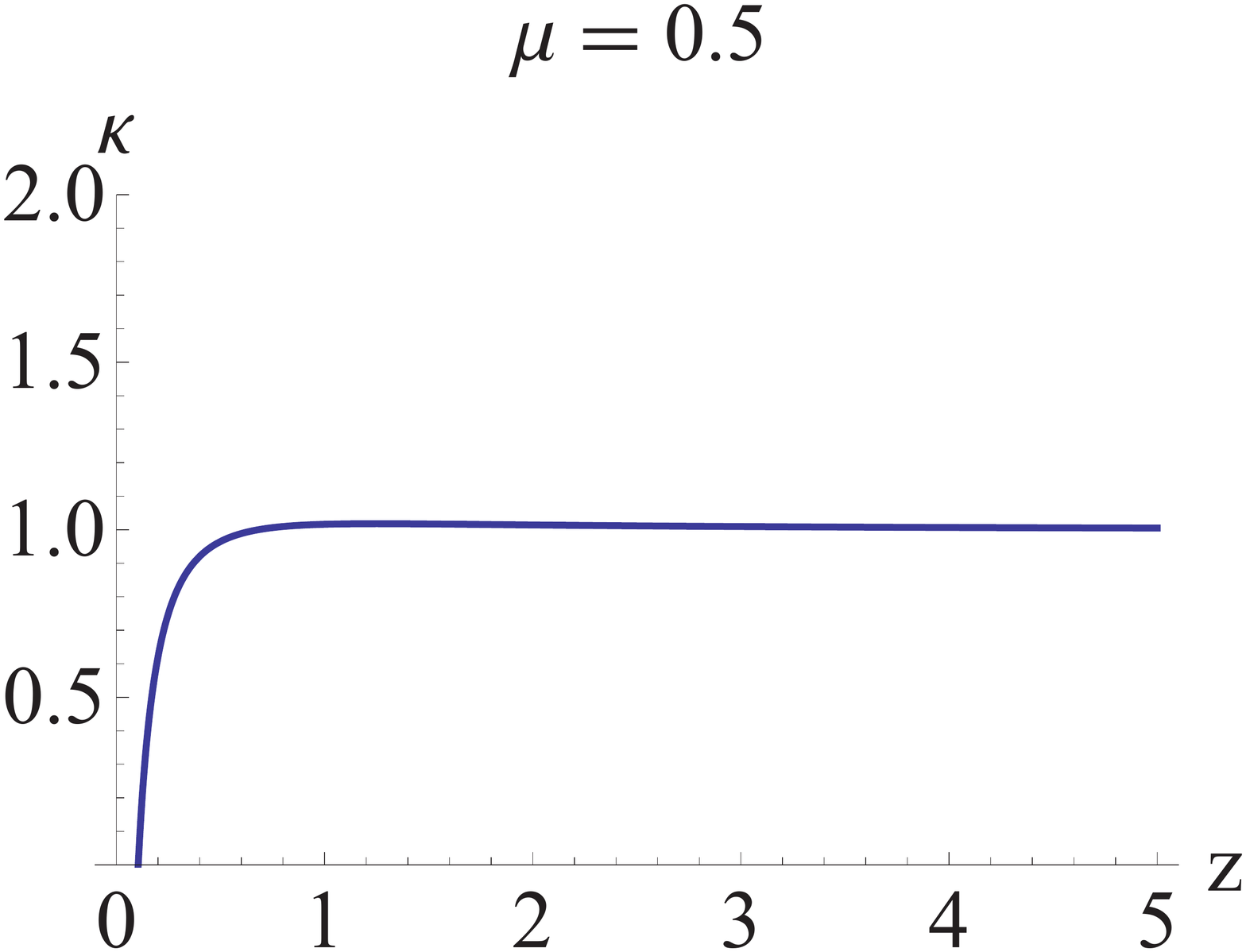}  
\end{center}
\caption[]{Dynamic surface gravity $\kappa(z)$ as described in (\ref{fullkappa}) for various values of $\mu$.}
\label{fig:kappa}
\end{figure}

We can calculate the dynamic surface gravity $\kappa$ using the definition (\ref{kappa}) proposed in \cite{Hayward99} or alternatively using the definition $\pm\kappa k_{\mu}=k^{\beta}\nabla_{[\mu}k_{\beta]}$, $\kappa\geq0$ proposed in \cite{Hayward97}. It should be noted that these two definitions only coincide on the trapping horizon \cite{Hayward97}. We find
\begin{align}
\kappa=\frac{1}{2}\epsilon^{\alpha\beta}\nabla_{\alpha}k_{\beta}=\frac{\sqrt{R+z^{1-\mu }}}{z}+\frac{(\mu -1) \sqrt{R+z^{1-\mu }} \left(R z^{\mu }+\mu  R z^{\mu }+2 z\right)}{4 \left(R z^{\mu }+z\right)^2}
\label{fullkappa}
\end{align}
where we have to insert (\ref{outertrapping}) for $R$ in order to obtain $\kappa$ on the outer trapping horizon. Some plots of $\kappa(z)$ are shown in figure \ref{fig:kappa} for representative values of $\mu\leq1$. The first thing that we should notice is that for $\mu=\pm1$ $\kappa$ is a constant in time and attains the correct values. For $\mu<-1$ we find that $\kappa$ is monotonously decreasing with $z$ and approaches the BTZ value $\kappa=1$ in the limit $z\rightarrow0$, while for $z\rightarrow\infty$ we find $\kappa\rightarrow+\infty$. For $|\mu|<1$ in contrast, we find $\kappa\rightarrow1$ for $z\rightarrow\infty$ while for small $\kappa$ a non-monotonous behaviour is possible. Starting from large values of $z$ and taking the limit $z\rightarrow0$ we find that at first $\kappa$ increases, only to attain a maximum for some $\kappa>0$ and then diverge to $-\infty$. In general, it is obvious that $\kappa$ attains the value $\kappa=1$ of the background metric in limits where the distortion $h_{\mu\nu}\sim 
z^{1-\mu}$ becomes small and $g_{\mu\nu}\approx\bg_{\mu\nu}$ whereas it shows a complicated behaviour where the distortion $h_{\mu\nu}$ is large. The values $z_0$ where $\kappa=0$ for $|\mu|<1$ are exactly the values where the outer trapping horizon switches from spacelike to timelike, as discussed in section \ref{sec:trappinghorizons}. This is another reason why one might doubt the validity of the trapping horizons as black hole boundary at least for small $z$ when $|\mu|<1$.

\begin{figure}[Hhtbp]                                 
\begin{center}
\includegraphics[width=0.32\linewidth]{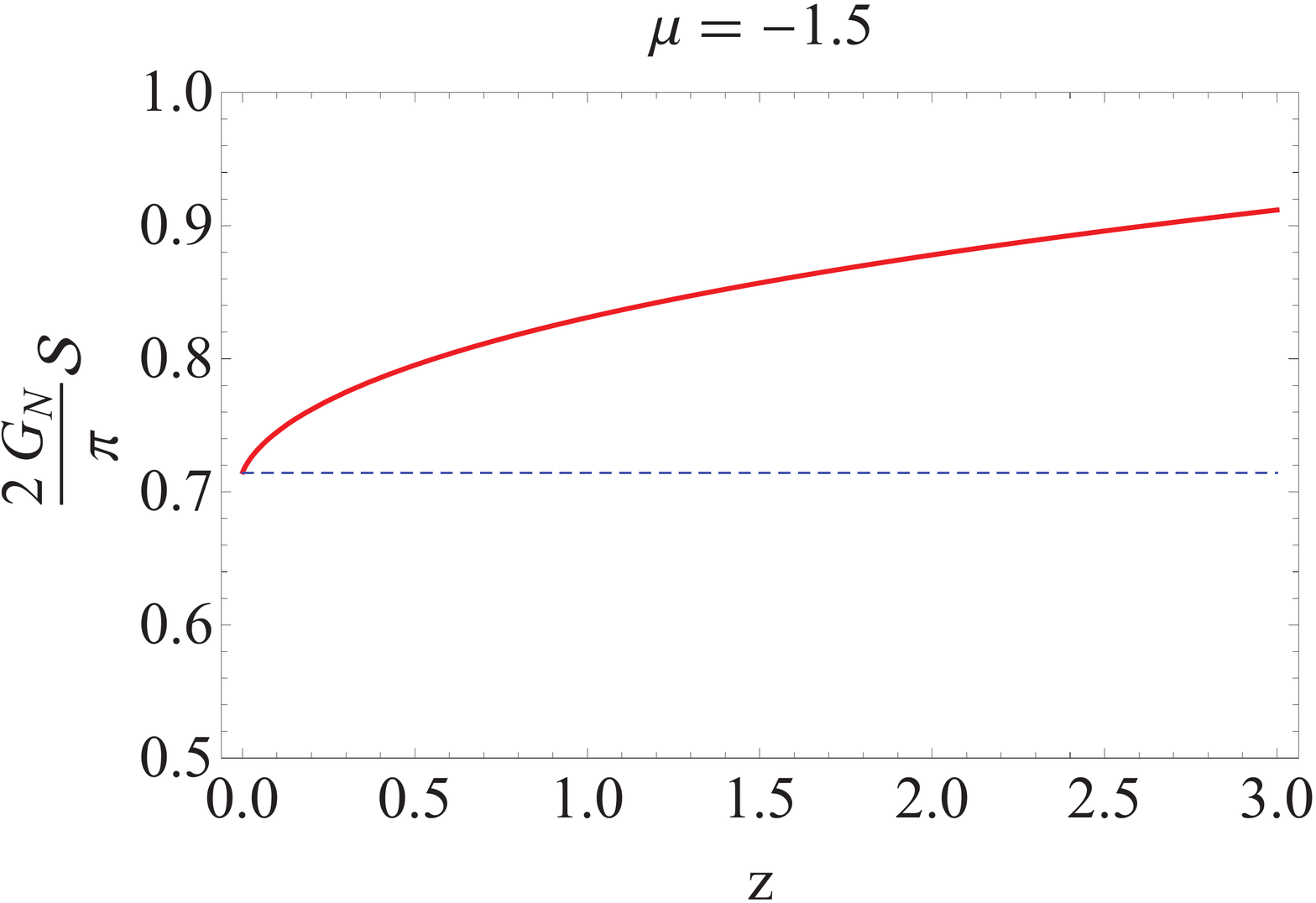}  
\includegraphics[width=0.32\linewidth]{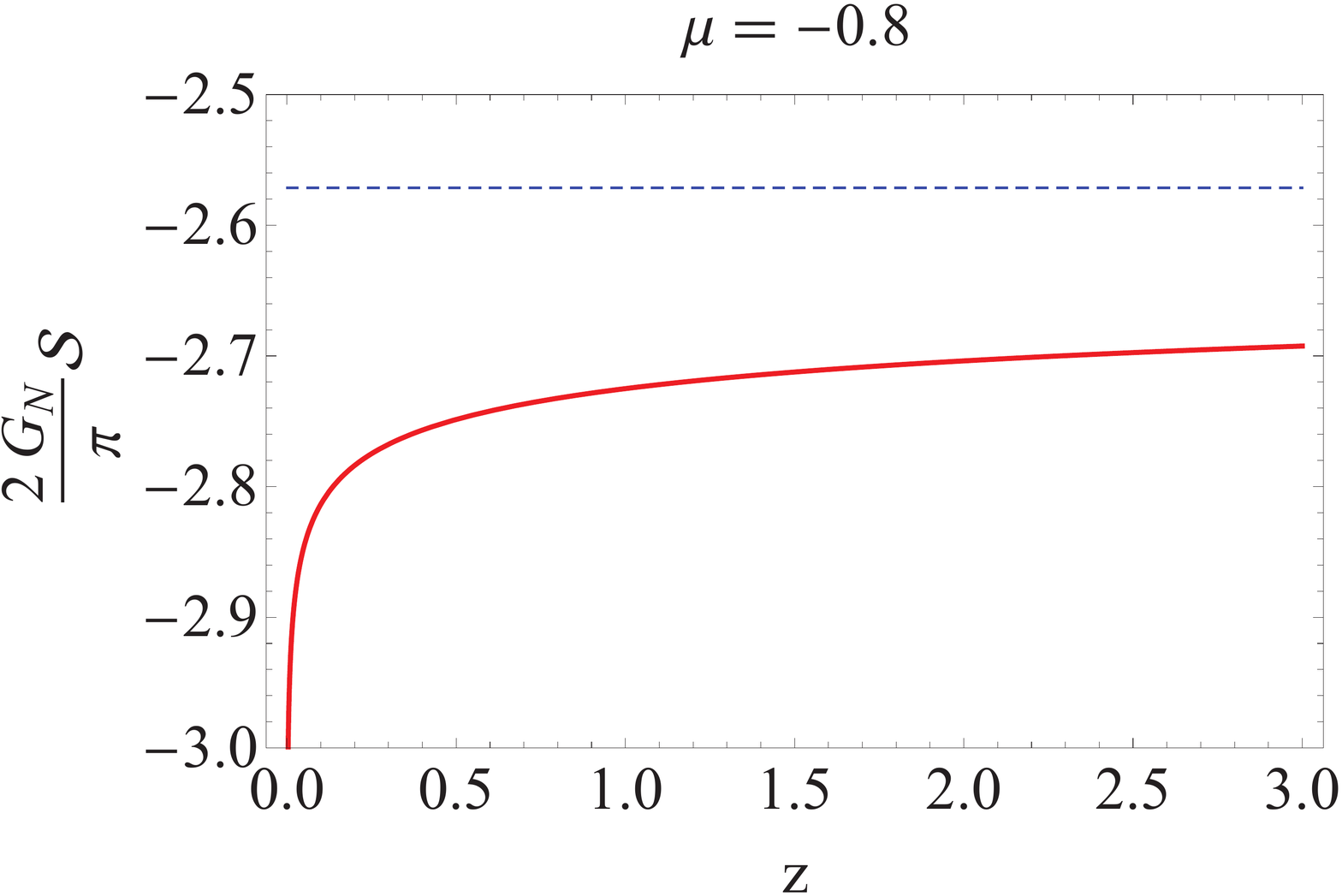}  
\includegraphics[width=0.32\linewidth]{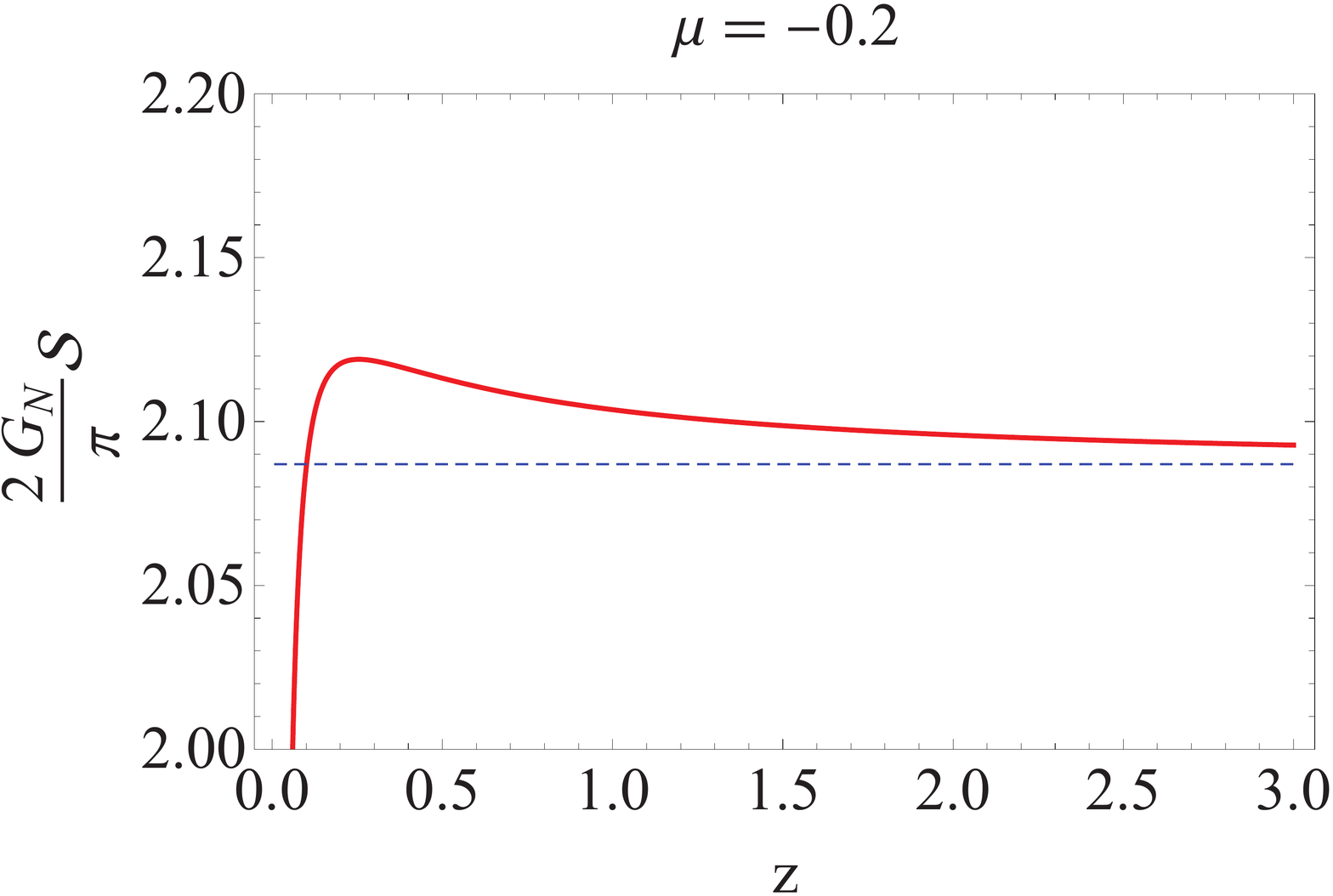}  
\includegraphics[width=0.32\linewidth]{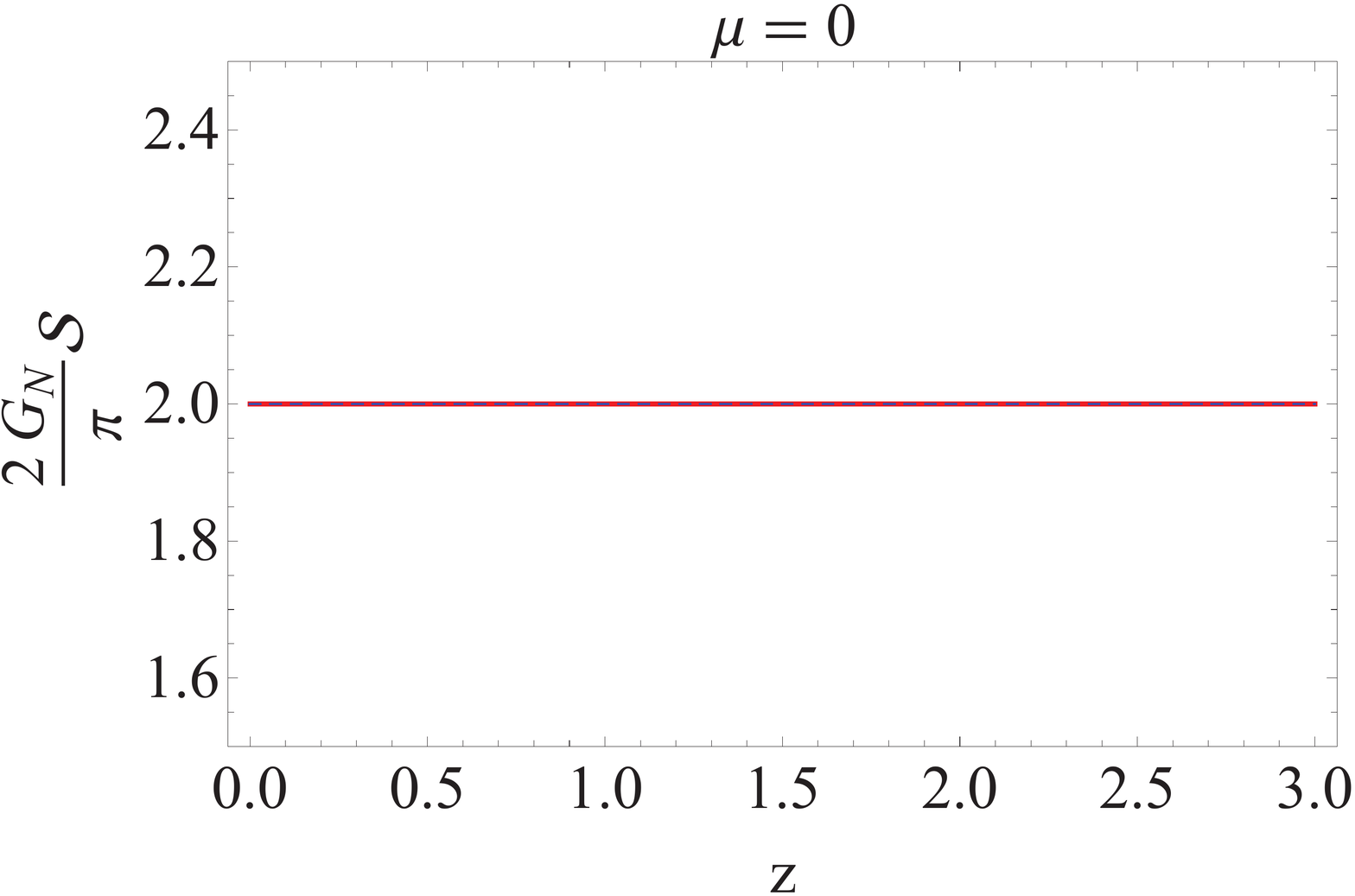}  
\includegraphics[width=0.32\linewidth]{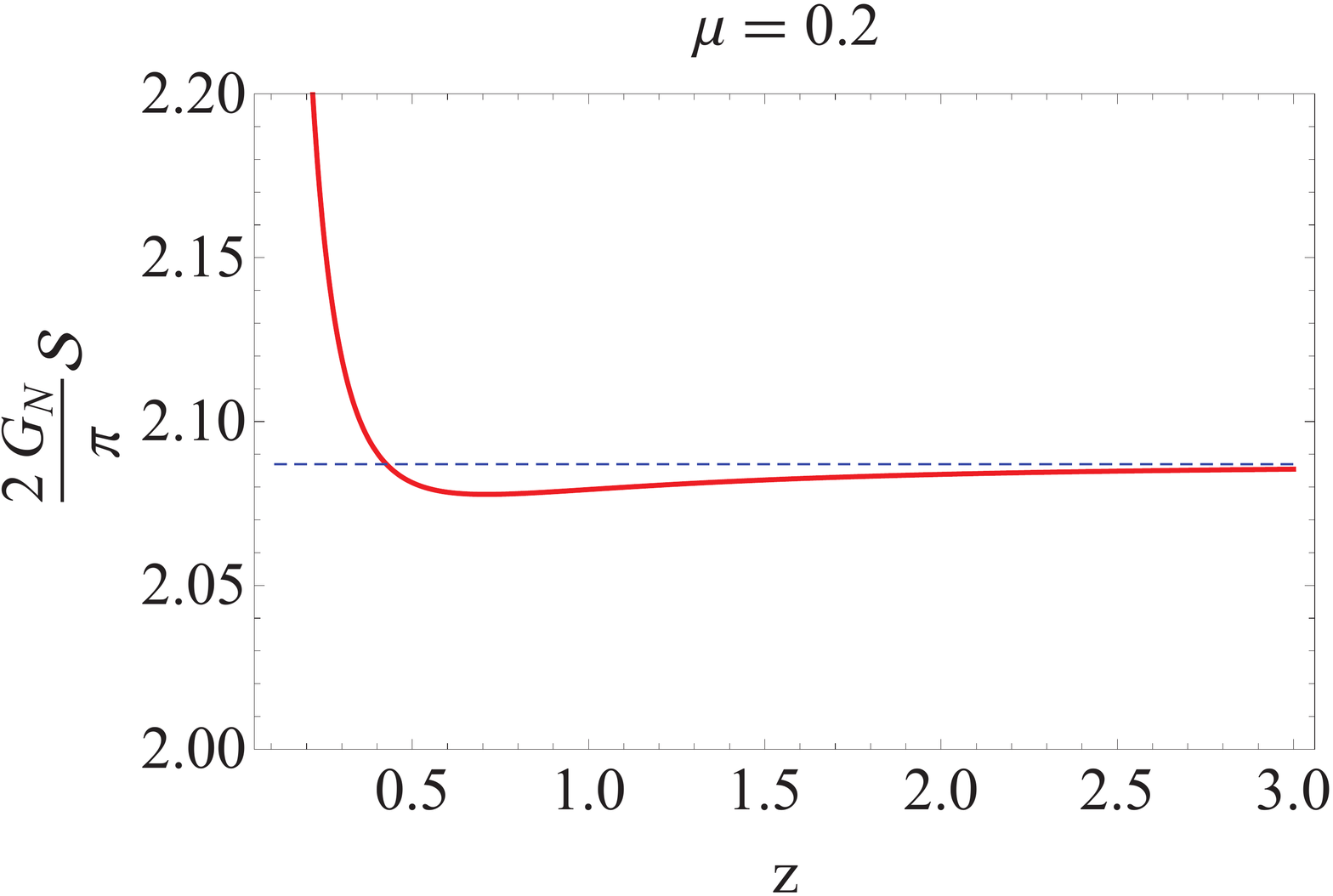}  
\includegraphics[width=0.32\linewidth]{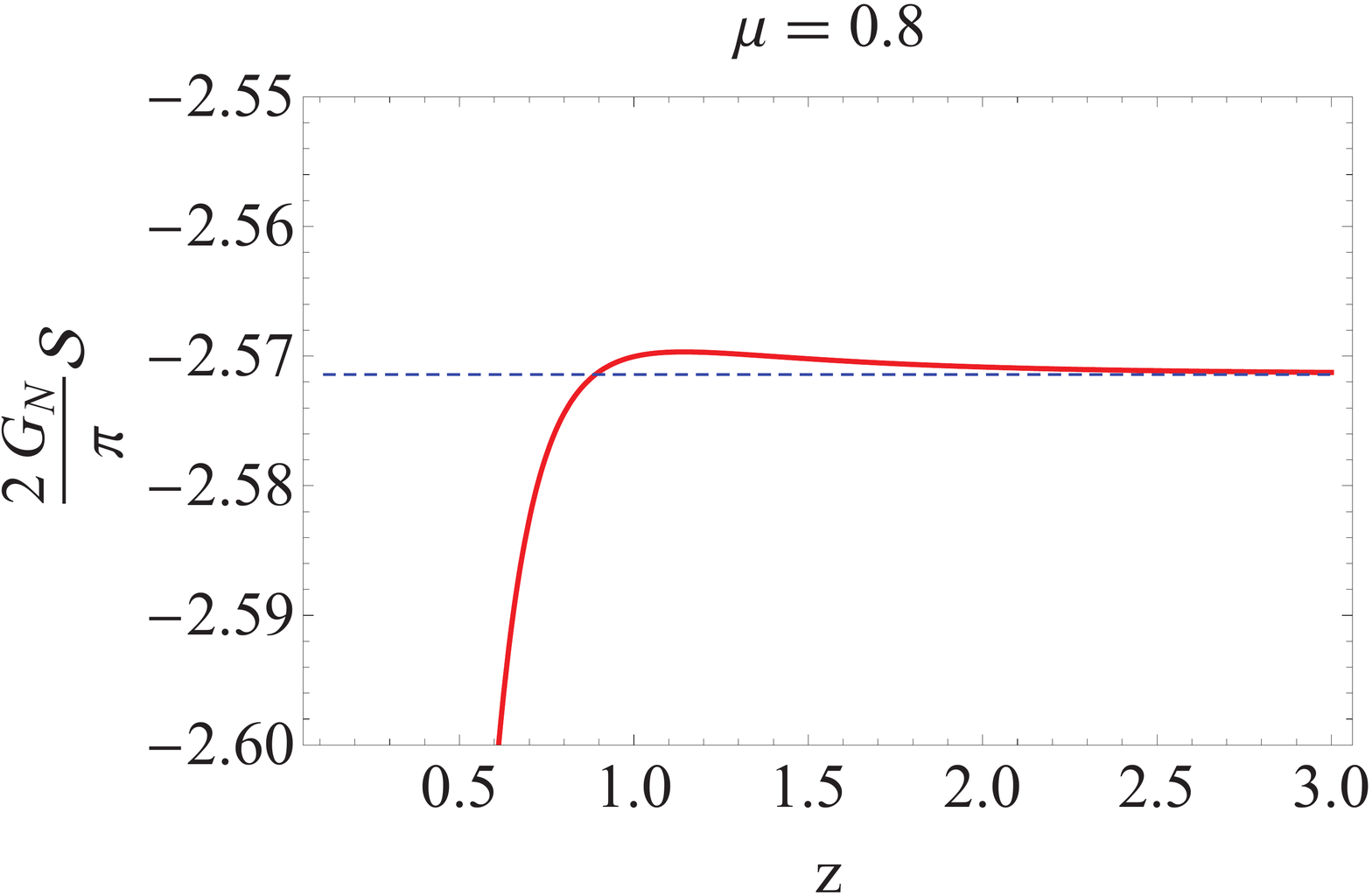}  
\end{center}
\caption[]{$\mathcal{S}(z)$ evaluated on the outer trapping horizon following Hayward's approach for different values of $\mu$, see (\ref{S_Hayward}). The dynamic entropy $\mathcal{S}(z)$ is shown as solid red line, the constant entropy value $\bar{\mathcal{S}}$ of the background metric $\bg_{\mu\nu}$ for the respective value of $m^2=\mu^2-\frac{1}{2}$ is shown as dashed blue line. %$\bar{\mathcal{S}}$ can be calculated from (\ref{S_NMG}) with $r_{+}=l=\sigma=1$.
}
\label{fig:S_NMG}
\end{figure}

We can now calculate the dynamic entropy according to Hayward's approach using (\ref{S_Hayward}). Some plots for $\mathcal{S}(z)$ for representative values of $\mu$ are given in figure \ref{fig:S_NMG}. In these plots, $z$ is used as a measure of time as the outer trapping horizons are monotonously increasing functions $R(z)$ at least for sufficiently large $z$, and as the coordinate $R$ can be used as a measure of time, see section \ref{sec:global coord}. Small values of $z$ will then correspond to the future, while large values of $z$ correspond to the past.

When discussing the results obtained for the dynamical entropy we should compare these to the values that the entropy $\bar{\mathcal{S}}$ of the background metric $\bg_{\mu\nu}$ would have as a function of $\mu$. For NMG we find with (\ref{S_Hayward}), $m^2=\mu^2-\frac{1}{2}$ (see section {\ref{sec:NMG}}) and $l=1$ that $\bar{\mathcal{S}}(\mu)=\frac{\sigma\pi}{2G_N}\left(1+\frac{1}{1-2 \mu ^2}\right)$ (see also \cite{Clement:2009gq}). 

For $\mu<-1$, $\mathcal{S}(z)$ is monotonously decreasing in time (i.e. increasing in z) for $\sigma=+1$ (see appendix \ref{sec:NMG0}) and approaching the value $\mathcal{S}=1$ for $z\rightarrow0$. The distortion $h_{\mu\nu}\sim z^{-1-\mu}$ becomes small in this limit and it is not surprising that $\mathcal{S}\rightarrow\bar{\mathcal{S}}$ as $g_{\mu\nu}\rightarrow\bg_{\mu\nu}$. Furthermore, with $m^2=\mu^2-\frac{1}{2}$ the limit $\mu\rightarrow-\infty$ corresponds to the limit where the NMG-action (\ref{NMGaction}) approaches the Einstein-Hilbert action, and thus the entropy becomes increasingly dominated by the horizon circumference which was shown to decrease with time in section \ref{sec:ev horizons}. Choosing $\sigma=-1$ as required by (\ref{ghostfree}) would result in an entropy $\mathcal{S}$ that is monotonously increasing from $-\infty$ for large $z$ to $-1$ for $z\rightarrow0$.

For $|\mu|<1$ the behaviour of $\mathcal{S}(z)$ is more complicated. First of all, it should be noted again that due to the unphysical behaviour of the trapping horizon discussed in section \ref{sec:trappinghorizons} the coordinate $z$ cannot be used as a time coordinate for arbitrarily small $z$. Above we saw that the surface gravity $\kappa$ vanishes at the $z$-value where the trapping horizon becomes timelike which leads to a divergence of $\mathcal{S}(z)$ at the same value of $z$. Secondly, for this range of $\mu$ $\mathcal{S}$ is generally not a monotonous function as can be seen in figure \ref{fig:S_NMG}. The behaviour for values $-1<\mu<1$ cannot be explained even qualitatively solely using the properties of NMG (such as unitarity, positivity of energy etc.) as the parameters of NMG, $\lambda$ and $m^2$, only depend on $\mu^2$ (see section \ref{sec:NMG}). This means that for example the qualitative differences in $\mathcal{S}(z)$ for $\mu=\pm0.2$ cannot be just due to properties of the action.

The value $\mu=0$ deserves special attention. For TMG this value has to be excluded due to the divergence in the action, but the metric (\ref{sachsmetric}) and the NMG action (\ref{NMGaction}) are well defined for this value. In this special case one finds $\mathcal{S}(z)=\bar{\mathcal{S}}=const.$ although the metric is clearly not stationary as can be seen from the surface gravity $\kappa(z)\neq const.$ or the time dependent circumference of the event horizon. Interestingly, $\mu=0$ corresponds to the special point $\frac{\lambda}{m^2}=1$ discussed in appendix \ref{sec:NMG0}.

\section{A note on the general 2+1 dimensional stationary Black Hole}
\label{sec:generalBH}

%We want to point out the significance of the results obtained in section \ref{sec:-1}. 
It was mentioned in section \ref{sec:-1} (and could of course be checked by straight forward calculations) that Hayward's approach reproduces the correct entropy for $\mu=-1$, $\Xi$ arbitrary, and we want to point out the significance of these results. The situation in three dimensions is special in the sense that only in this case the Kodama vector can be defined for rotating black holes, as only for $d=3$ the axial symmetry of a rotating black hole equals the symmetry of a $(d-2)$-sphere used in section \ref{sec:Kodama and Entropy}. The fact that for $\mu=-1$ Hayward's approach reproduces correct values for surface gravity and entropy although the Kodama vector is not a Killing vector seems at first nontrivial. Instead of now applying Hayward's approach to other (ideally rotating) black hole solutions known in NMG (see e.g. \cite{moreonNMG,Clement:2009gq,Clement:2009ka,Oliva:2009ip,AyonBeato:2009nh}) one after the other, we can look at the general stationary black hole metric in $2+1$ dimensions. 

Assume that there is a metric which allows for two commuting Killing vectors $\eta$ (timelike) and $\chi$ (spacelike). One can then find a coordinate system in which the coordinates are $\tilde{t},\rho$ and $\tilde{\phi}$ such that $\eta^{\alpha}\partial_{\alpha}=\partial_{\tilde{t}}$ and $\chi^{\alpha}\partial_{\alpha}=\partial_{\tilde{\phi}}$. Assume furthermore that in these coordinates $\tilde{t}\in]-\infty,+\infty[$ and $\tilde{\phi}\in[0,2\pi[$ with $\tilde{\phi}\sim\tilde{\phi}+2\tilde{\pi}$. This means that $\tilde{\phi}$ is an angular coordinate. One can now always perform a coordinate transformation that yields a metric   
\begin{align}
g_{\mu\nu}=
\left(
\begin{array}{ccc}
g_1(r) & 0 & g_2(r) \\
 0 & g_3(r)^{-1} & 0 \\
g_2(r) & 0  & r^2  \\
\end{array}
\right)
\label{generalBH}
\end{align}
in coordinates $t,r,\phi$ with $\partial_{t}=\partial_{\tilde{t}}$  and $\partial_{\phi}=\partial_{\tilde{\phi}}$. In this metric, the binormal can equivalently be defined as $\epsilon^{\mu\nu}=\frac{1}{r\sqrt{-g}}\epsilon^{\mu\nu\lambda}$ where $\epsilon^{\mu\nu\lambda}$ denotes the ordinary Levi-Civita-Symbol with values $-1$, $0$ or $1$. Using this it is easy to show that the Kodama vector reads
\begin{align}
k^{\mu}=\frac{r}{\sqrt{-g}}\left(\delta^{\mu}_{1}-\frac{g_2(r)}{r^2}\delta^{\mu}_{3}\right)
\nonumber
\end{align}
It is noteworthy that $k^{\mu}k_{\mu}=-g_3(r)$ which means that this metric only describes a genuine black hole with a (trapping) horizon if a coordinate singularity is present in the Schwarzschild-like coordinates. The horizon is then defined by the radial coordinate $r_{+}$ with $g_3(r_{+})=0$\footnote{As the metric is stationary, we assume that trapping and event horizon agree.}. The determinant of (\ref{generalBH}) reads $g=\frac{r^2g_1(r)-g_2(r)^2}{g_3(r)}$. As $r=r_{+}$ is only supposed to be a coordinate singularity, we assume $g$ to be well defined there, which means $r_{+}^2g_1(r_{+})-g_2(r_{+})^2=0$ (see also \cite{Jacobson:2007tj} for a related issue). If this assumption is true, there exists a Killing vector 
\begin{align}
\xi^{\mu}=\left(\delta^{\mu}_{1}-\frac{g_2(r_{+})}{r_{+}^2}\delta^{\mu}_{3}\right)\neq k^{\mu}
\label{KillingVektor}
\end{align}
which is null on the horizon, and this is exactly the Killing vector used to calculate the black hole entropy according to \cite{Wald48,Kang49,Wald50}. It is known that this vector vanishes on the bifurcation surface of the black hole \cite{Wald48} and the same is obviously true for the Kodama vector which on the horizon is just $k^{\mu}=const.\cdot\xi^{\mu}$. Therefore, as the entropy can be evaluated on the bifurcation surface \cite{Wald48}, the term proportional to $k^{\mu}$ can be neglected in (\ref{Qcomponents}) for the black hole (\ref{generalBH}). As one can show that $\nabla_{[\alpha}k_{\beta]}|_{r=r_{+}}=const.\cdot\epsilon_{\alpha\beta}|_{r=r_{+}}$, and as this constant is canceled from the integral (\ref{S_Hayward}) by the prefactor $\kappa^{-1}$, Hayward's approach yields the entropy
\begin{align}
\mathcal{S}=\frac{-2\pi}{16\pi G_N}\int_{\Sigma}X^{\alpha\beta\gamma\delta}\epsilon_{\alpha\beta}\epsilon_{\gamma\delta}\sqrt{\gamma}d\phi
\end{align}
which is also obtained from the usual ansatz using the Killing vector (\ref{KillingVektor}) \cite{Kang49,Wald50}.

This proves that Hayward's approach reproduces the correct entropy for the general stationary (but possibly rotating) black hole in the framework of an arbitrary $2+1$ dimensional covariant theory of gravity\footnote{It should be noted that in the sense of \cite{Wald48,Kang49,Wald50}, TMG is not a covariant theory. This has for example been pointed out in \cite{Tachikawa,Bonora}} when one can use an expression of the form (\ref{Qcomponents}). There is a little subtlety here: It has already been noted in \cite{Kang49} that for a general covariant theory, $Q^{\mu\nu}$ might depend on arbitrary high derivatives $\nabla_{\alpha}...\nabla_{\mu}k_{\nu}$ of the used vector field (using this yields $S_1$ in (7) of \cite{Kang49}). This expression can then always be brought into the form (\ref{Qcomponents}) using identities that hold if $k_{\mu}$ is a Killing vector, yielding $S_2$ in (7) of \cite{Kang49}. While in \cite{Wu} it was proposed to use the full Noether potential (i.e. $S_1$ in \cite{Kang49}) for Hayward's 
approach, in \cite{Hayward99b} it was suggested to use $S_2$. As shown, at least in the latter case Hayward's approach reproduces the correct entropy for stationary black holes in $2+1$ dimensions.

%==============================================
%=============  Conclusion  =========================
%==============================================
\section{Conclusion}
\label{sec:Conclusion}

We investigated the metric (\ref{sachsmetric}) and showed that for general $\mu$ it describes a dynamical black hole with inner and outer event and trapping horizons. The metric is a solution of NMG for suitable parameters $m^2(\mu)$ and $\lambda(\mu)$, and reduces to previously known stationary BTZ black holes for $\mu=\pm1$. We applied the three dimensional Kodama vector and Hayward's approach to dynamical black hole entropy to our dynamic black hole metric (\ref{sachsmetric}). For $|\mu|\neq1$ the results are in apparent contradiction with the second law. For $\mu=-1$ where due to the emergence of the additional parameter $\Xi$ the metric (\ref{sachsmetric}) describes a whole family of \textit{rotating} BTZ black holes, the correct entropy is reproduced although Kodama and Killing vector do not agree. In fact we proved that this is the case for the general stationary but possibly rotating black hole in $2+1$ dimensions. In appendix \ref{sec:IW} we will apply the definition proposed by Iyer and Wald in \
cite{Wald50} to the dynamical black holes (\ref{sachsmetric}).
\\
\\

\noindent {\Large \bf  Acknowledgments}\\ \\
M. F. would like to thank Olaf Hohm for many useful discussions. I. S. would like to thank the Center for the Fundamental Laws of Nature at Harvard University for hospitality during the initial stages of this project. This project was supported in parts by the DFG Transregional Collaborative Research Centre TRR 33,
the DFG cluster of excellence ``Origin and Structure of the Universe'' as well as the DAAD project  54446342. 

%=============================================================================================================
%============================         Appendix        ======================================================================
%=============================================================================================================

%\part{Appendix}

\begin{appendix}

%==============================================
%==========       NMG       ===========================
%==============================================
\section{New Massive Gravity}
\label{sec:NMG0}

The action of NMG can be written in the form\footnote{Unfortunately, there seem to be competing conventions on how to present this action in the literature. The form employed in \cite{NMG,moreonNMG,correlators} has the integrand $\sigma'R-2\lambda'm'^2+\frac{1}{m'^2}K$. The dictionary for comparing results obtained with the two actions reads: $\sigma=\sigma'$, $\lambda=\lambda'm'^2/\sigma'$ or $\lambda'=-\lambda/m^2$ and $\sigma'm'^2=-m^2$}
 \cite{moreonNMG,correlators,Chakhad:2009em}
\begin{align}
  S_{NMG}=\frac{\sigma}{16\pi G_N}\int d^{3}x\sqrt{-g}\left(R-2\lambda-\frac{1}{m^2}K\right)
\label{NMGaction}
\end{align}  
where $\lambda$ is the cosmological constant, $\sigma=\pm1$ is the overall sign of the action that is irrelevant for the equations of motion but relevant for conserved charges, and $K=R_{\mu\nu}R^{\mu\nu}-\frac{3}{8}R^2$ is the trace of the tensor \cite{NMG}
\begin{align}
K_{\mu\nu}=\ &2\nabla^2 R_{\mu\nu}-\frac{1}{2}\left(\nabla_{\mu}\nabla_{\nu} R+g_{\mu\nu}\nabla^2 R\right)-8R_{\mu}{}^{\alpha}R_{\alpha\nu}+\frac{9}{2}RR_{\mu\nu}
\label{KofNMG}
\\
&+\left(3R^{\mu\nu}R_{\mu\nu}-\frac{13}{8}R^2\right)g_{\mu\nu}
\nonumber
\end{align}
It should be noted that the parameter $m^2$ will be allowed to have positive as well as negative values \cite{moreonNMG}. The equations of motion read \cite{NMG,Chakhad:2009em}
\begin{align}
R_{\mu\nu}-\frac{1}{2} g_{\mu\nu} R +\lambda g_{\mu\nu}-\frac{1}{2m^2}K_{\mu\nu}=0
\label{EOMofNMG}
\end{align}
and taking the trace obviously yields 
\begin{align}
R-6\lambda+\frac{1}{m^2}K=0
\label{TraceofNMG}
\end{align}
This means that in contrast to Einstein-Hilbert gravity and TMG, in NMG the Ricci scalar $R$ is not fixed by the cosmological constant. For a maximally symmetric spacetime (such as $\text{AdS}_3$) with $R_{\mu\nu}=2\Lambda g_{\mu\nu}$ and therefore $R=6\Lambda$ the expressions containing $\nabla$ in (\ref{KofNMG}) will automatically vanish yielding $K_{\mu\nu}=-\frac{1}{2}\Lambda^2 g_{\mu\nu}$ and consequently $K=-\frac{3}{2}\Lambda^2$. Upon inserting these expressions, the equations of motion (\ref{EOMofNMG}) reduce to
\begin{align}
-\Lambda g_{\mu\nu} +\lambda g_{\mu\nu}+\frac{\Lambda^2}{4m^2}g_{\mu\nu}=0
\nonumber
\end{align}
Evidently, for a maximally symmetric spacetime with curvature $\Lambda$ to be a solution of NMG the parameters need to fulfill\footnote{For our conventions of signs and prefactors in (\ref{NMGaction}), this is equivalent to the condition presented in (2) of \cite{correlators} and in (1.11) of \cite{moreonNMG} which relates the AdS-radius $l$ ($\Lambda=-\frac{1}{l^2}$) of possible AdS solutions of NMG to the parameters of the theory.} \cite{moreonNMG}
\begin{align}
\Lambda=2m^2\left(1\pm\sqrt{1-\frac{\lambda}{m^2}}\right)
\label{AdSinNMG}
\end{align}
For $\frac{\lambda}{m^2}>1$, maximally symmetric solutions are obviously not possible.

When the theory is linearized around a maximally symmetric background metric satisfying $R_{\mu\nu}=2\Lambda g_{\mu\nu}$ (for example $\text{AdS}_3$), it can be proven \cite{moreonNMG} that NMG is ghost-free when the condition 
\begin{align}
\frac{m^2}{\sigma}\left(\Lambda+2m^2\right)<0
\label{ghostfree}
\end{align}
is satisfied. Together with (\ref{AdSinNMG}) and the Breitenloher-Freedman bound \cite{moreonNMG}%\todo{cite Sungmin} 
\begin{align}
2m^2\geq\Lambda
\label{BF}
\end{align}
there are several inequalities that restrict the physically acceptable sets of parameters $\sigma$, $\lambda$ and $m^2$ for which linearization about an AdS background yields a unitary, ghost free theory \cite{moreonNMG}. %We will come back to these issues in section \ref{sec:NMG_sol}. 

NMG has two propagating bulk degrees of freedom corresponding to massive graviton modes with spin $\pm2$, except for $-\frac{\lambda}{m^2}=-1$ or $-\frac{\lambda}{m^2}=3$ and $\Lambda=-2m^2$ \cite{moreonNMG}. In the first exceptional case there appears a so-called single partially massless mode \cite{moreonNMG}. The second exceptional case, where $-\frac{\lambda}{m^2}=3$, was shown to be a very special situation. There, the linearized Lagrangian equals the Proca Lagrangian for a spin 1 field with squared mass $8m^2$\cite{moreonNMG}. As in this case unitarity requires $m^2\sigma<0$, the spin 1 modes are Tachyons for $\sigma=1$ but physical for $\sigma=-1$ \cite{moreonNMG}.

When the parameters of NMG are chosen in order to allow AdS-vacua with $R_{\mu\nu}=-\frac{2}{l^2} g_{\mu\nu}$ ($l>0$), then a dual CFT can be conjectured to exist according to the $\text{AdS}_3$/$\text{CFT}_2$-correspondence, having left- and right-moving central charges \cite{moreonNMG}
\begin{align}
c_L=c_R=c=\frac{3l\sigma}{2G_N}\left(1-\frac{1}{2m^2l^2}\right)
\label{c_NMG}
\end{align}
The sign of the central charges obviously depends on $\sigma$ and changes when $m^2=\frac{1}{2l^2}$. Positivity of the central charge is required as well for unitarity of the CFT as for positivity of entropy and mass of the BTZ black hole \cite{moreonNMG}. Unfortunately, as realized in \cite{moreonNMG} the conditions on the parameter space arising from the requirement $c\geq0$ are inconsistent with the requirements arising from the desire to have unitary positive-energy modes apart from the special case $-\frac{\lambda}{m^2}=3$ where $c=0$. %As noted in \cite{moreonNMG} this situation is quite similar to the one discussed for TMG in the previous subsection. 

%=========================================================================================
%============================       Iyer Wald         =================================================
%=========================================================================================

\section{Iyer-Wald approach to Dynamic Entropy}
\label{sec:IW}

%============================        Idea         =================================================
\subsection{Idea}
Immediately after the discovery that black hole entropy can be calculated via the Noether charge approach in \cite{Wald48} ideas were presented in \cite{Wald48,Kang49,Wald50} how these results could be used to generalize the definition of black hole entropy to the non-stationary case. In this section, we will make use of the prescription for defining dynamical black hole entropy that was put forward by Vivek Iyer and Robert Wald in \cite{Wald50}, and which we will call the \textit{Iyer-Wald approach}.

The entropy of a black hole can be calculated by an integral of the form \cite{Wald50}
\begin{align}
\mathcal{S}(\Sigma')=2\pi\int_{\Sigma'}X^{\gamma\delta}\epsilon'_{\gamma\delta}
\label{IW103}
\end{align}
where $\Sigma'$ is a spacelike slice of the horizon and $\epsilon'$ is the binormal to $\Sigma'$. It was shown in \cite{Kang49} that in the stationary case the value of (\ref{IW103}) is independent of the choice of the slice $\Sigma'$ and that we can consequently choose $\Sigma'$ to be the bifurcation surface $\Sigma$. In the dynamic case the entropy will be a function of time by definition. Thus, if an expression of the form (\ref{IW103}) is still valid in the dynamical case, the choice of spacelike slice $\Sigma'$ corresponds to the choice of time at which the entropy is to be computed. What is now needed for a definition of dynamical black hole entropy is a generalization of the integrand $X^{\gamma\delta}$ to the dynamical case \cite{Wald50}.

The Iyer-Wald approach is based on the following idea \cite{Wald50}: Consider a spacetime with metric $g_{\mu\nu}$ with a dynamical outer event horizon, and take a spacelike slice $\Sigma'$ of this horizon corresponding to a certain time. Then apply a transformation $g_{\mu\nu}\rightarrow\tilde{g}_{\mu\nu}$ that generates an entirely new metric in which the horizon slice $\Sigma'$ is embedded as the bifurcation surface of a stationary black hole. The entropy $\tilde{\mathcal{S}}(\Sigma')$ of this black hole can readily be calculated using the appropriate formula for the stationary case (\ref{IW103}) and is set to be equal to the dynamic black hole entropy $\mathcal{S}(\Sigma')$. This embedding of the horizon slice does obviously not change the horizon area. Therefore, for dynamical black holes in Einstein-Hilbert gravity the entropy calculated using the Iyer-Wald approach is proportional to the horizon surface. Due to the area theorem \cite{Frolov,HE} this means that for Einstein-Hilbert gravity a second law 
can be inferred for the dynamic entropy following from the Iyer-Wald approach \cite{Wald50}.

In the following, we will give the definition of the transformation $g_{\mu\nu}\rightarrow\tilde{g}_{\mu\nu}$ which Wald and Iyer proposed in \cite{Wald50} in order to calculate dynamic black hole entropy.

\begin{definition}\cite{Wald50}:
Let $\Sigma'$ be a $(d-2)$ dimensional spacelike surface with a field $M^{\alpha_1,\alpha_2,...}{}_{\beta_1,\beta_2,...}$ defined on it. $M^{\alpha_1,...}{}_{\beta_1,...}$ will be called \textit{boost invariant on $\Sigma'$} if for every point $\mathcal{P}$ on $\Sigma'$, $M^{\alpha_1,...}{}_{\beta_1,...}$ is invariant under Lorentz boosts in the tangent space at $\mathcal{P}$ in the $(1+1)$ dimensional plane orthogonal to $\Sigma'$. When at the point $\mathcal{P}$ one chooses a set of orthogonal spacelike vectors $s_{i}^{\mu}$ ($i\in\{1,...d-2\}$) tangent to $\Sigma$ and $l^{\mu}$ and $n^{\mu}$ as independent null vectors orthogonal to $\Sigma'$, then these vectors can be used to define a tetrad $e^{\mu}{}_{a}$\footnote{Technically, the term \textit{tetrad} is only for $d=4$. The general term is \textit{frame field} or \textit{vielbein}.}. One can then expand $M$ in this basis:
\begin{align}
M^{\alpha_1,\alpha_2,...}{}_{\beta_1,\beta_2,...}=\tilde{M}^{a_1a_2,...}{}_{b_1,b_2,...}e^{\alpha_1}{}_{a_1}e^{\alpha_2}{}_{a_2}e_{\beta_1}{}^{b_1}e_{\beta_2}{}^{b_2}...
\label{tetexpansion}
\end{align}
The tensor $M$ is boost invariant if and only if the basis expansion coefficients $\tilde{M}^{a_1,...}{}_{b_1,...}$ are only non-vanishing for terms with equal numbers of $l^{\mu}$'s and $n^{\mu}$'s \cite{Wald50}.
\end{definition}
In order to illustrate this definition and obtain an important result, we will now prove for $d=3$ that the metric is always boost invariant on $\Sigma'$ \cite{Wald50}: Let us choose the tetrad  $e^{\mu}{}_{0}=l^{\mu}$, $e^{\mu}{}_{1}=n^{\mu}$ and $e^{\mu}{}_{2}=s^{\mu}$. The tetrad expansion (\ref{tetexpansion}) of the metric can easily be found as the relation $\eta_{mn}=g_{\mu\nu}e^{\mu}{}_{m}e^{\nu}{}_{n}$ holds \cite{Wald}. This relation defines the three dimensional Minkowski metric in lightcone coordinates, as we are working with a null tetrad. We can read off $\eta_{00}=l_{\mu}l^{\nu}=0$, $\eta_{11}=n_{\mu}n^{\nu}=0$ and $\eta_{02}=l_{\mu}s^{\mu}=0=\eta_{20}=\eta_{12}=\eta_{21}$ due to orthogonality. Therefore, the inverse relation $g_{\mu\nu}=\eta_{mn}e_{\mu}{}^{m}e_{\nu}{}^{n}$ yields the expression $g_{\mu\nu}=\eta_{01}l_{\mu}n_{\nu}+\eta_{10}l_{\nu}n_{\mu}+\eta_{22}s_{\mu}s_{\nu}$ where in each term the number of $l^{\mu}$'s equals the number of $n^{\mu}$'s. Therefore, the metric is always boost 
invariant on $\Sigma'$.

\begin{definition}\cite{Wald50}:
Let $\Sigma'$, $M^{\alpha_1,...}{}_{\beta_1,...}$ and the tetrad be defined as in the previous definition. When $M^{\alpha_1,...}{}_{\beta_1,...}$ is not boost invariant, then we can extract the \textit{boost invariant part} $\widehat{M^{\alpha_1,...}{}_{\beta_1,...}}$ of $M^{\alpha_1,...}{}_{\beta_1,...}$ by defining it to be the field on $\Sigma'$ that is obtained when in the tetrad expansion (\ref{tetexpansion}) only the terms with equal numbers of $l^{\mu}$'s and $n^{\mu}$'s are kept.
\end{definition}

It should be noted that this definition of the boost invariant part is independent of the choice of the tetrad \cite{Wald50}. Although the metric itself is always boost invariant this does not hold for objects containing derivatives of the metric, such as Christoffel symbols and curvature tensors. Hence, it is useful to define a metric $g^{I}_{\mu\nu}$ which is boost invariant and also yields boost invariant curvature tensors \cite{Wald50}. In order to achieve this goal, Iyer and Wald proposed to define a certain coordinate system in the neighbourhood of $\Sigma'$ in the following approach \cite{Wald50}\footnote{For simplicity, we will restrict the discussion to three dimensions in the following.}: On $\Sigma'$ we define again a null-tetrad with vectors $l^{\mu}$, $n^{\mu}$ and $s^{\mu}$ such as in the definitions above. Furthermore, we require the normalization $l_{\mu}n^{\mu}=-1$. The neighbourhood around $\Sigma'$ that we are going to investigate is assumed to be small enough that every point $\mathcal{P}'
$ lies on a unique geodesic orthogonal to $\Sigma'$. This geodesic is assumed to be (affinely) parametrized in such a way that $\mathcal{P}'$ is at unit affine distance from $\Sigma'$, and $\gamma^{\mu}$ is then assumed to be the tangent of the geodesic at the intersection point $\mathcal{P}$ with $\Sigma'$. The coordinates of $\mathcal{P}'$ are now defined to be $U$, $V$ and $s$ where $U$ and $V$ are the components of $\gamma^{\mu}$ along $l^{\mu}$ and $n^{\mu}$ respectively and $s$ is the coordinate of $\mathcal{P}$ on $\Sigma'$.

In these coordinates the Taylor expansion of the metric $g_{\mu\nu}$ around $\Sigma'$ (being defined by $U=0=V$, $s$ being arbitrary) reads \cite{Wald50}:
\begin{align}
g_{\alpha\beta}=\sum_{n,m=0}^{\infty} \frac{U^{m}V^{n}}{m!n!}\left(\frac{\partial^{m+n}g_{\alpha\beta}(U,V,s)}{\partial U^m \partial V^n}\right)\Bigg|_{U=V=0}
\nonumber
\end{align}
In an arbitrary coordinate system this equation reads
    \begin{align}
    g_{ab}=\sum_{n,m=0}^{\infty} \frac{U^{m}V^{n}}{m!n!}
  \left( l^{c_1}\cdots l^{c_m}n^{c_{m+1}}\cdots n^{c_{m+n}}\partial_{c_1}\cdots\partial_{c_{m+n}}g_{ab}\right)\big|_{U=V=0}
    \label{IW109}
    \end{align}
where $U$ and $V$ are to be understood as implicit functions of the new coordinates. It should be noted that in our three dimensional case the term $\left( l^{c_1}\cdots\partial_{c_{m+n}}g_{ab}\right)\big|_{U=V=0}$ is a constant as $U$ and $V$ are set to zero and as the metric does not depend on the remaining angular coordinate\footnote{For simplicity we always use slices of the horizon which are generated by the Killing vector $\partial_\phi$.}.

Wald and Iyer proposed \cite{Wald50} to define a new metric $g_{\mu\nu}^{I_{q}}$ by truncating the infinite series in (\ref{IW109}) at the level $n+m=q$ and replacing each of the expressions $\partial_{c_1}\cdots g_{\alpha\beta}$ by its boost invariant part. They realized \cite{Wald50} that the metric $g_{\mu\nu}^{I_{\infty}}$ has a Killing vector field $\xi=U\partial_U-V\partial_V$ which vanishes on the slice $\Sigma'$ which is defined by $U$=$V$=0. Thus, this Killing vector field generates a Killing horizon with $\Sigma'$ as bifurcation surface. The idea of Wald and Iyer to define dynamical black hole entropy with respect to a horizon slice $\Sigma'$ was to construct the metric tensor $g_{\mu\nu}^{I_{q}}$ with $q$ being larger than the highest derivative order appearing in the entropy formula and calculate the entropy of this new metric using the appropriate formula for the stationary case \cite{Wald50}. 

%============================       Calculation         =================================================
\subsection{Calculation}

In order to apply the method described in the previous subsection it seems that we have to find the exact coordinate transformation $U=U(z,R)$, $V=V(z,R)$, $s=y+s'(z,R)$\footnote{We assume a coordinate transformation that respects the Killing symmetry generated by $\partial_y$, in the sense that $\partial_y=\partial_s$.} that allows us to write the metric (\ref{dsforzyR}) with respect to these coordinates. However, for TMG there is an easier way to do this calculation.  

For stationary black holes in TMG Tachikawa \cite{Tachikawa} found that the contribution of the Chern-Simons term to the entropy reads\footnote{As mentioned above, the contribution from the Einstein-Hilbert term will still be proportional to the circumference of the horizon slice.}
\begin{align}
\mathcal{S}_{CS}(\Sigma')=\frac{1}{8G_{N}\mu}\int_{\Sigma'}\epsilon_{\alpha\beta}g^{\alpha\nu}g^{\beta\mu}\Gamma_{\mu\nu\rho}dx^{\rho}
\label{Tachikawa}
\end{align}
where $\epsilon_{\alpha\beta}$ denotes the binormal as defined in section \ref{sec:Kodama and Entropy}. For the non-stationary case, according to Wald and Iyer one would have to calculate the Christoffel symbols $\Gamma_{\mu\nu\rho}(g^{I}_{\alpha\beta})$ with respect to the new metric. The construction of (\ref{IW109}) is based on the substitution of the expressions $\partial_{c_1}\cdots g_{ab}$ by their boost invariant parts. Hence, one can ask if there is the possibility to calculate the boost invariant part of $\Gamma_{\mu\nu\rho}(g_{\alpha\beta})$ instead of $\Gamma_{\mu\nu\rho}(g^{I}_{\alpha\beta})$. For more general theories such as NMG we can furthermore ask whether instead of calculating for example the Ricci scalar $R(g^{I}_{\alpha\beta})$ we can write the Ricci scalar as a function of the metric and it's derivatives ($R(g_{\alpha\beta},\partial_{c}g_{\alpha\beta},\partial_{d}\partial_{c}g_{\alpha\beta})$) and subsequently substitute these expressions by their boost invariant parts. As we will see 
this is only possible for expressions with at most first derivative order of the metric.

As the metric is boost invariant it is obvious from (\ref{IW109}) that on the horizon ($U=V=0$)
\begin{align}
g^{I}_{ab}\big|_{\Sigma'}=\widehat{g_{ab}}\big|_{\Sigma'}=g_{ab}\big|_{\Sigma'}
\nonumber
\end{align}
In addition, for the first derivative we find $\partial_y g^{I}_{ab}\big|_{\Sigma'}=0=\partial_y g_{ab}\big|_{\Sigma'}$ due to symmetry, and for $\partial_c g^{I}_{ab}\big|_{\Sigma'}$ with $c\neq y$:
\begin{align}
\partial_c g^{I}_{ab}\big|_{\Sigma'}&=\left(\sum_{n,m=0}^{\infty} 
\left[m\frac{U^{m-1}V^{n}}{m!n!}\partial_{c}U+n\frac{U^{m}V^{n-1}}{m!n!}\partial_{c}V\right]
  \left( l^{c_1}\cdots\partial_{c_{m+n}}g_{ab}\right)\bigg|_{\Sigma'}\right)\Bigg|_{\Sigma'}
\nonumber
\\
&=\left[\partial_{c}U l^{c_1}+\partial_{c}V n^{c_1}\right]\big|_{\Sigma'}
  \left(\widehat{\partial_{c_1}g_{ab}}\right)\Big|_{\Sigma'}
\nonumber
\\
&=\delta_{c}^{c_1}
  \left(\widehat{\partial_{c_1}g_{ab}}\right)\Big|_{\Sigma'}
\nonumber
\\
&=\widehat{\partial_{c}g_{ab}}\big|_{\Sigma'}
\nonumber
\end{align}
In this derivation we used that $\partial_U=l^{\alpha}\partial_{\alpha}$ and $\partial_V=n^{\alpha}\partial_{\alpha}$. From the coordinate relations it then follows that
\begin{align}
\partial_c=\left(\frac{\partial U}{\partial x^c}\right)\partial_U+\left(\frac{\partial V}{\partial x^c}\right)\partial_V+\left(\frac{\partial s}{\partial x^c}\right)\partial_s
\nonumber
\\
\Rightarrow\left((\partial_{c}U)l^{\alpha}+(\partial_{c}V)n^{\alpha}\right)\partial_{\alpha}
=\left(\delta_{c}^{\alpha}-(\partial_{c}s)\delta_s^{\alpha}\right)\partial_{\alpha}
\nonumber
\end{align}
Here the term containing $\delta_s^{\alpha}$ can be omitted as the derivative of the metric with respect to the angular coordinate vanishes due to $\partial_s$ being a Killing vector. It is therefore justified to substitute $\partial_{c}U 
l^{c_1}+\partial_{c}V n^{c_1}$ by $\delta_{c}^{c_1}$ in the above derivation.

Using the same approach one can show that
\begin{align}
\partial_d\partial_c g^{I}_{ab}\big|_{\Sigma'}\neq\widehat{\partial_{d}\partial_{c}g_{ab}}\big|_{\Sigma'}
\nonumber
\end{align}
due to terms involving expressions such as $(\partial_{d}\partial_{c}U)\big|_{\Sigma'}(l^{c_1}\widehat{\partial_{c_1}g_{ab}})\big|_{\Sigma'}$ that are not vanishing and that cannot be eliminated in a way similar to the one used above.

Therefore, we can calculate the dynamic entropy according to Iyer and Wald without knowing the exact coordinate transformation to the coordinate system $U$, $V$, $s$ for TMG, but not for NMG where higher derivatives of the metric are needed.

%============================        Discussion         =================================================
\subsection{Discussion}
\label{sec:IWdiscussion}

Let us now discuss the results for TMG. The first consistency check of our calculations is that for $\mu=\pm1$ we know (see the footnote in section \ref{sec:-1}) that we need to find $\mathcal{S}_{TMG}(\Sigma')=\frac{\pi}{2G_N}$. This is indeed the case, but in some sense this is trivially the case for an unfortunate reason: While we have $\epsilon^{\mu\nu}\widehat{\Gamma_{\mu\nu\rho}}\neq\epsilon^{\mu\nu}\Gamma_{\mu\nu\rho}$ in general, we obtain $\epsilon^{\mu\nu}\widehat{\Gamma_{\mu\nu y}}=\epsilon^{\mu\nu}\Gamma_{\mu\nu y}$ which is the only part of the integrand that matters, as on the horizon $\int_{\Sigma'}(...)dx^\rho=\int_{0}^{2\pi}(...)\big|_{z=z',R=R'}dy$ in (\ref{Tachikawa}). This means that taking the boost invariant part does not give other results than the direct use of (\ref{Tachikawa}) would have given. 

\begin{figure}[Hhtb]                                 
\begin{center}
\includegraphics[width=0.4\linewidth]{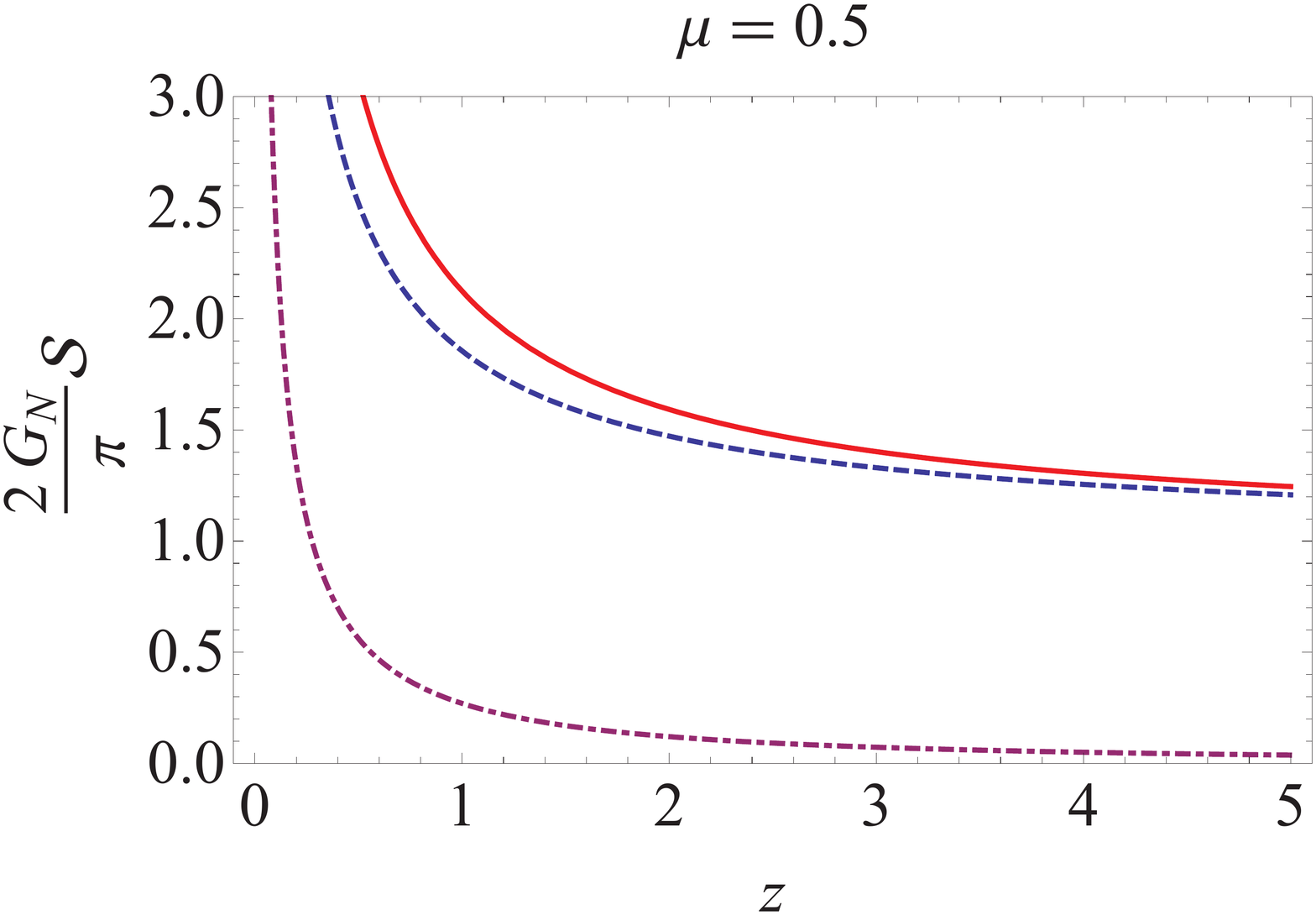}  
\includegraphics[width=0.4\linewidth]{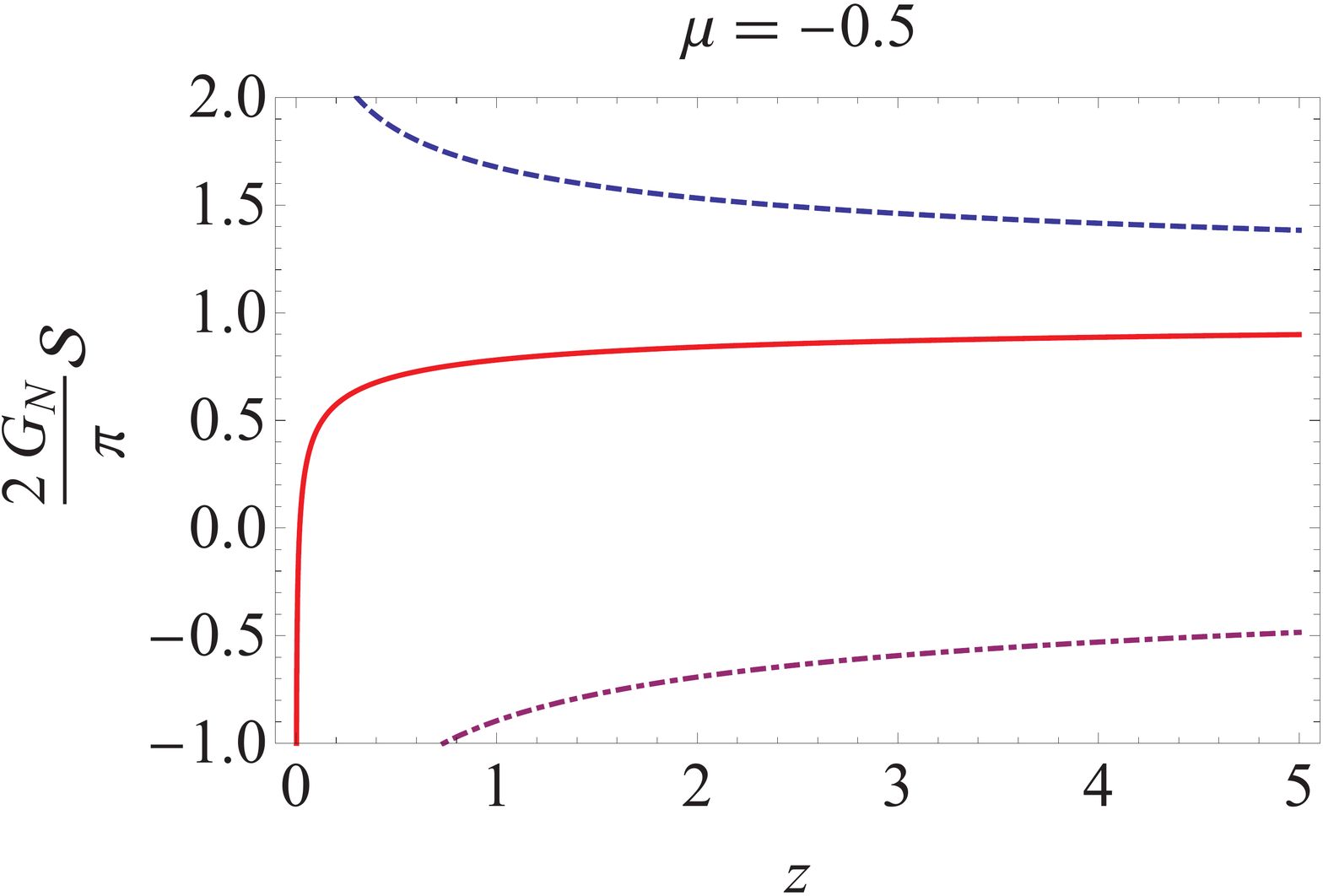}  
\includegraphics[width=0.4\linewidth]{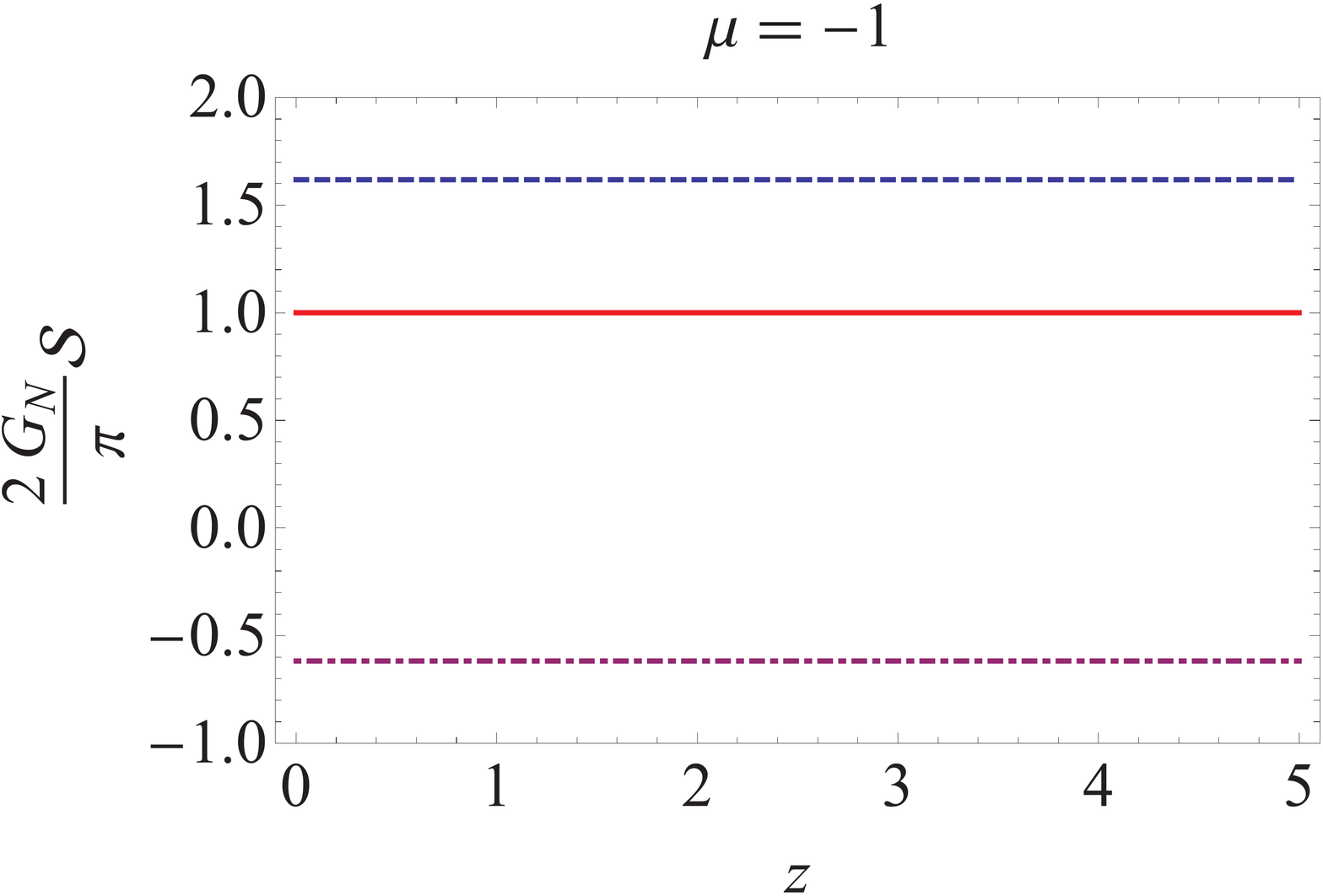}  
\includegraphics[width=0.4\linewidth]{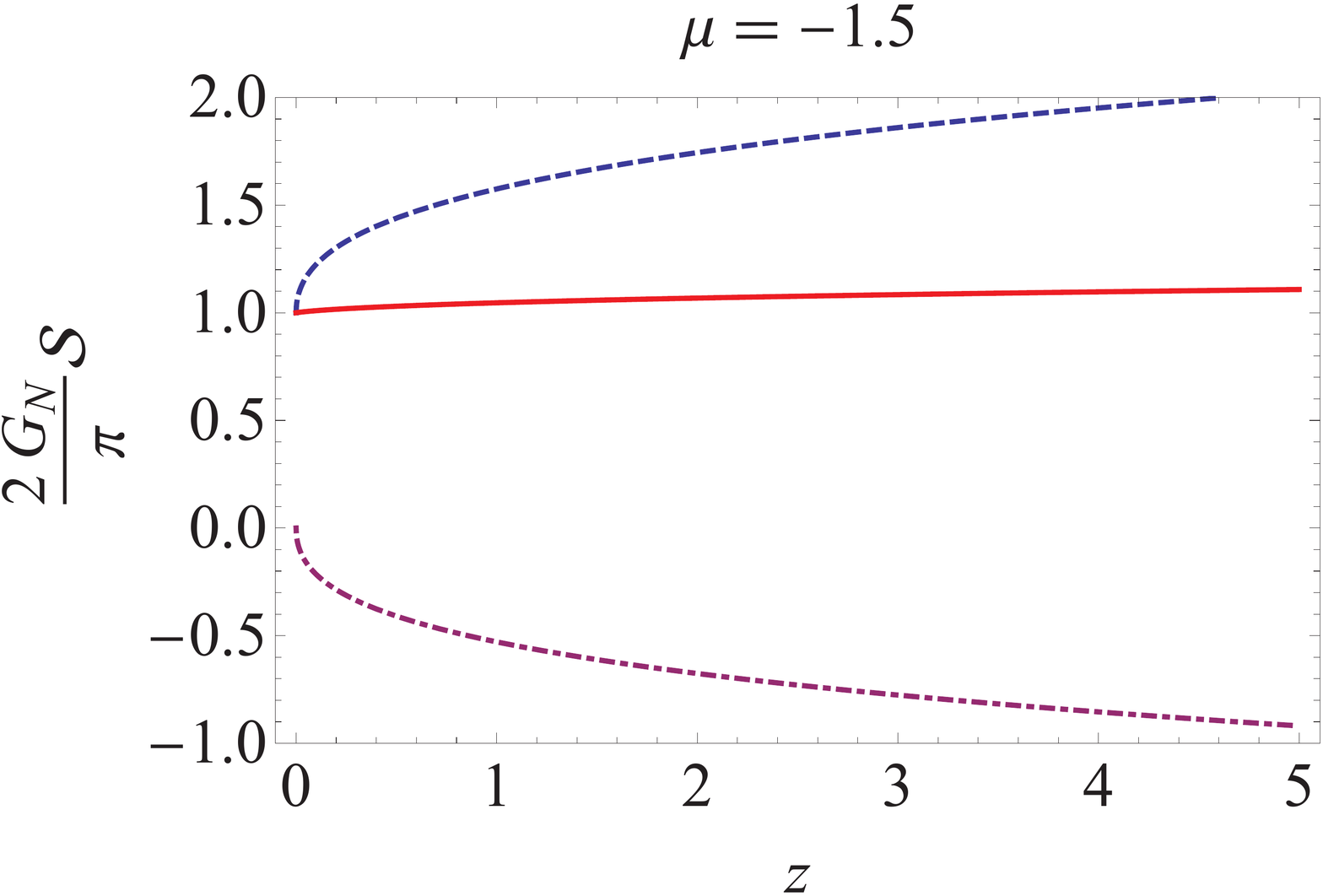}  
\end{center}
\caption[]{$\mathcal{S}(z)$ in the framework of TMG evaluated on the outer event horizon following the Iyer-Wald approach for different values of $\mu$. The dynamic entropy $\mathcal{S}(z)$ is shown as solid red line, the contribution from the Einstein-Hilbert term of the action (proportional to the horizon circumference) is shown as dashed blue line, the contribution from the Chern-Simons term (\ref{Tachikawa}) is shown as dot-dashed purple line.}
\label{fig:S_TMG}
\end{figure}

For $\mu<1$ and $\mu\neq-1$ we find that the dynamic entropy $\mathcal{S}_{TMG}(\Sigma')$ will not be constant. As in section \ref{sec:Kodama and Entropy} it will be easiest to take spacelike slices of spacetime denoted by a certain value of $R\equiv R'$ which leads for the intersection with the horizon also to a certain value of $z\equiv z'$. As we wrote the horizons as functions $R(z)$ in section \ref{sec:ev horizons} for the event horizon and in section \ref{sec:trappinghorizons} for the trapping horizon, we can therefore also write the dynamic entropy as a function $\mathcal{S}(z)$. Due to monotonicity of the event horizons and for large enough $z$ also of the trapping horizons, smaller values of $z$ will correspond to the future and larger values of $z$ will correspond to the past. Plots of the results for $\mathcal{S}(z)$ for several $\mu\leq1$ can be found in figure \ref{fig:S_TMG}. We find that when evaluated on the event horizon, $\mathcal{S}_{TMG}(z)$ is increasing (and actually diverging) in time 
for $\mu>0$ and decreasing in time for $\mu<0$, where as $z\rightarrow0$ it diverges to $-\infty$ for $-1<\mu<0$ and limits to $\frac{\pi}{2G_N}$ for $\mu\leq-1$\footnote{As the event horizon can only be studied numerically for $\mu\neq\pm1$ there is always the risk that a certain behaviour at some limit is due to numerical problems.}. As expected, there is always a limit in which the entropy approaches the value $\mathcal{S}_{TMG}(z)\rightarrow\frac{\pi}{2G_N}$ which is the same limit in which the distortion $h_{\mu\nu}$ becomes small, i.e. $z\rightarrow+\infty$ for $|\mu|<1$ and $z\rightarrow0$ for $\mu<-1$.

The great advantage of the Iyer-Wald approach is that it is not intrinsically limited to slices of the event horizon. Indeed, there have been arguments that in the dynamic cases entropy should in fact be assigned to the trapping horizon rather than to the event horizon, see \cite{Hayward99,Hayward99b,Wu} (section \ref{sec:Kodama and Entropy}) and \cite{Hubeny0,Hubeny1} for two different approaches to dynamic black hole entropy that both favour trapping or apparent horizons over event horizons. We can therefore in our calculations substitute the event horizon (\ref{outerevhorizon}) with the trapping horizon (\ref{outertrapping}) and calculate the dynamic entropy with respect to this quantity. It should be noted that for $|\mu|<1$ this might be problematic for small values of $z$ due to the unphysical behaviour of the trapping horizon discussed in section \ref{sec:trappinghorizons}. Therefore, in our results for $\mathcal{S}_{TMG}(z)$ the variable $z$ cannot be interpreted as a time variable anymore. As it 
turns out, the qualitative behaviour of $\mathcal{S}(z)$ calculated with respect to the trapping horizons is not different from the qualitative behaviour of the entropy when calculated with respect to the event horizon. 

The results obtained using the Iyer-Wald approach are clearly not satisfactory, as they indicate a decreasing entropy as a function of time for some parameters $\mu$. This might be due to either the method we used for calculating the entropy or to the properties of TMG. On one hand, it was already pointed out in a note added to \cite{Wald50} that the entropy calculated using the Iyer-Wald approach is not invariant under field redefinitions, in contrast to what should be expected for physical reasons. On the other hand, it was discussed in \cite{D1} that TMG has some unphysical properties for $l\mu\neq\pm1$, making a possible violation of the second law of black hole thermodynamics less surprising.   

\end{appendix}

%altmodisches Literaturverzeichnis:
\providecommand{\href}[2]{#2}\begingroup\raggedright\endgroup

%neumodisches Literaturverzeichnis:-
%\bibliography{masterpaper}{}
%\bibliographystyle{JHEP}

\end{document}